\documentclass[pre,twocolumn,aps,groupaddress]{revtex4}
\usepackage[T1]{fontenc}
\usepackage[utf8]{inputenc}
\usepackage{array}
\usepackage{booktabs}
\usepackage{mathrsfs}
\usepackage{mathbbol}
\usepackage{multirow}
\usepackage{amsmath}
\usepackage{amsthm}
\usepackage{amssymb}
\usepackage{natbib}
\usepackage{stmaryrd}
\usepackage{graphicx}
\usepackage{latexsym}
\usepackage{textcomp}
\usepackage{mathtools}
\usepackage{xcolor}
\usepackage[normalem]{ulem}
\usepackage{soul}
\usepackage{changes}
\usepackage{adjustbox}
\usepackage[citecolor=magenta,colorlinks=true]{hyperref}
\usepackage{lineno}
\usepackage{stackrel}
\usepackage[toc,page,titletoc]{appendix}
\usepackage{bm}
\usepackage{comment}
\newcommand{\Pe}{\text{Pe}}
\newcommand{\peclet}{\text{P\'eclet~}}

\definecolor{mycolor1}{rgb}{0.1, 0.6, 0.6}
\begin{document}
\title{Traveling waves, phase separation and pattern formation in\\ the active stepping stone model}
\author{Rashmiranjan Bhutia}
\address{Tata Institute of Fundamental Research, Hyderabad, 500046, India}
\author{Prasad Perlekar}
\address{Tata Institute of Fundamental Research, Hyderabad, 500046, India}
\author{Kabir Ramola}
\address{Tata Institute of Fundamental Research, Hyderabad, 500046, India}
\date{\today}
\begin{abstract}
Selective advantage leads to the proliferation of a species into an ecosystem. On the other hand, self-propulsion can lead to dynamic clustering of individuals. To analyze the intricate effects produced by such competing processes, we introduce a minimal model for a proliferating active population, where run-and-tumble migration is combined with a stochastic birth-death process. Beginning with the microscopic dynamics of the particles, we derive the corresponding hydrodynamic equations and validate them through numerical simulations. We demonstrate that the inclusion of activity leads to the emergence of novel morphological patterns, which are sensitive to the specific rules governing migration. Furthermore, we show that activity significantly alters the characteristics of invasion waves, increasing their propagation speed. However, beyond a threshold, excessive activity disrupts the wave-like nature of the invasion process. We also extend our study to two dimensions and analyze the instabilities that lead to pattern formation in such systems. 
\end{abstract}
\maketitle

Active matter systems ranging from avian flocks to bacterial suspensions operate far from thermodynamic equilibrium, exhibiting rich collective behaviors~\cite{marchetti2013hydrodynamics,cavagna2010scale,te2026colloquium,toner1995long,jain2024inertia,huang2024active}. While their physical properties have been extensively characterized~\cite{liu2019self,sanchez2012spontaneous,narayan2007long}, their evolutionary dynamics and demographic fluctuations remain relatively less explored~{\cite{hallatschek2023proliferating,cates2010arrested,curatolo2020cooperative}. Bacterial populations represent a prototypical system where motility and proliferation unfold on comparable timescales~\cite{hallatschek2023proliferating,korolev2010genetic}, a coupling capable of generating complex spatiotemporal structures such as phase separation and intricate morphological patterns~\cite{cates2010arrested,curatolo2020cooperative,ben1997snowflake}. In passive systems and spatial population genetics, proliferating spatial dynamics are classically modeled through Branching Brownian Motion (BBM)~\cite{ramola2014universal,ramola2015spatial,brunet2013genealogies,brunet2008universal}, the stepping-stone framework~\cite{kimura1964stepping,duty2000broken,korolev2010genetic}, and reaction-diffusion models governed by Fisher-Kolmogorov-Petrovsky-Piscounov (FKPP) equations~\cite{kolmogorov2019studies,van1988front,doering2003interacting,brunet1997shift,hallatschek2009fisher}. However, existing descriptions combining persistent active motility with birth-death dynamics are predominantly phenomenological~\cite{cates2010arrested,curatolo2020cooperative,toner2012birth}, lacking a microscopic foundation that unifies active particle dynamics with branching and evolutionary models.

In this Letter, we introduce a microscopic lattice model that systematically connects active lattice gases exhibiting motility induced phase separation (MIPS)~\cite{kourbane2018exact,erignoux2016hydrodynamic} to the stepping-stone model of evolutionary dynamics and BBM-like demographic processes~\cite{kimura1964stepping,duty2000broken,korolev2010genetic,ramola2014universal,ramola2015spatial}. From these microscopic rules, standard coarse-graining techniques~\cite{lefevre2007dynamics,martin1973statistical} yield hydrodynamic equations that closely match particle-level Monte Carlo simulations, providing an agent-based foundation that unifies activity, genealogy, and spatial demographics. In the continuum limit, the model reduces to an active field theory~\cite{wittkowski2014scalar} with density-dependent mobility and time-reversal-symmetry-breaking demographic terms. Solving the coupled hydrodynamic equations yields a generalized phase diagram that recovers known morphologies~\cite{cates2010arrested,curatolo2020cooperative,meissner2024introduction,yadav2025coarsening} while also producing additional disordered, phase-separated, traveling-wave, as well as microphase patterns consistent with biological observations~\cite{budrene1991complex,budrene1995dynamics,curatolo2020cooperative}. We also study the nature of invasion (FKPP) waves in this system and show that activity enhances a single pulled front at low activity, however above a threshold it destabilizes and distorts the interface, in contrast with passive FKPP fronts~\cite{brunet1997shift,hallatschek2009fisher}.

\begin{figure}[t!]
\centering
\includegraphics[width=1\linewidth]{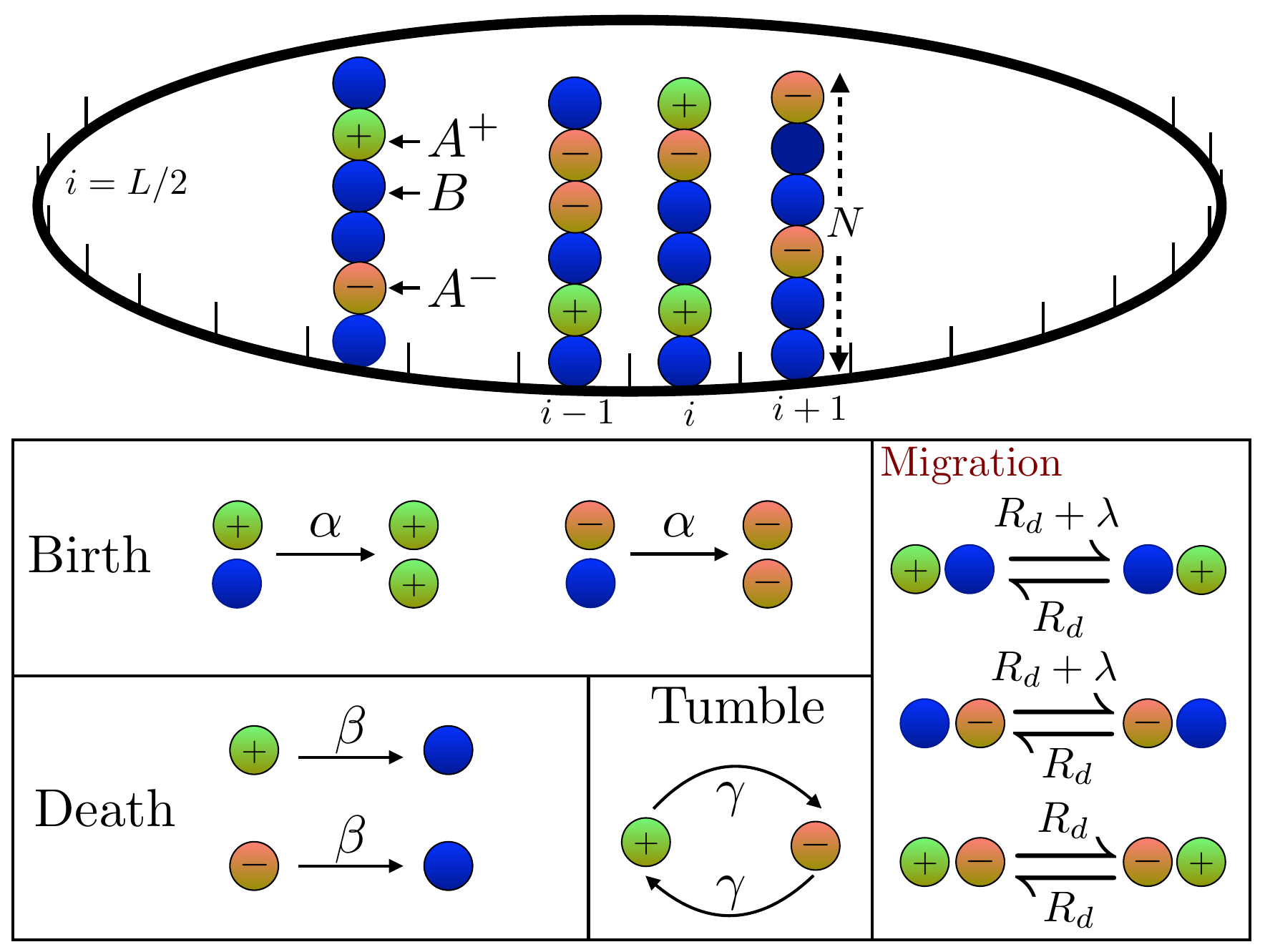}
\caption{Schematic of the active stepping-stone model in one dimension. Lattice sites $i$ contain active individuals $A$ (red/green) with a flippable directional spin $\sigma = \pm 1$ (right/left) and passive individuals $B$ (blue). The dynamics of the system include random neighbor exchanges across sites $i$ and $i+1$ and local on-site reactions.}
\label{fig: SS_Model}
\end{figure}

{\it The active stepping stone model (ASM)}: 
\label{sec:model_def}
We describe the one-dimensional version of the active stepping-stone model (with analogous generalizations in higher dimensions~\cite{SI}). The system is a periodic lattice of length ${l}=La$ on $\left(-l/2,~l/2\right]$ with sites $i=-L/2+1,...,-1,0,1,..., L/2$ separated by $a$. Each site (deme) contains a fixed total population $N$ of two alleles, $A$ and $B$~(see Fig.~\ref{fig: SS_Model}). The two-allele setting captures genetic drift and fixation~\cite{korolev2010genetic,doering2003interacting,pigolotti2013growth,singha2020fixation,bhutia2026diffusive}. Fixed $N$ is motivated by constant nutrient availability~\cite{Moran, wright1931evolution,duty2000broken}. The allele of type $A$ is additionally imbued with an activity, which we model as preferential motion in the positive or negative direction: $A^+$ and $A^-$, respectively. The migration rules are inspired by active lattice-gas models, which are amenable to hydrodynamic descriptions~\cite{kourbane2018exact} as well as a characterization of their fluctuations~\cite{agranov2021exact,jose2023current}.

At each time step: {\bf (1)} from a randomly chosen site, two of the $N$ individuals are selected and undergo
(a) birth: $A^{\pm}$+$B$ $\rightarrow$ $A^{\pm}$+$A^{\pm}$ at rate $\alpha$. The offspring inherit the parent's polarity~\cite{soares2019polarity,takamura2025spatially}, 
(b) death: $A^{\pm}$ $\rightarrow$ $B$ at rate $\beta$,
\label{eq:reaction_rules}
{\bf (2)} tumble: spins flip with rate $\gamma$, and {\bf (3)} individuals migrate between neighboring demes. The $+$ particles are biased to move to the right [Fig.~\ref{fig: SS_Model}] with a rate $\lambda$; additionally, we allow purely diffusive exchanges between all neighboring particles with rate $R_d$~\cite{kourbane2018exact}. Generalizing the rules to higher dimensions is straightforward (SM~\ref{sec:ASS_2d}~\cite{SI}), with the active directions of the particles aligned along the lattice directions~\cite{jose2022active,jose2022first}.

The dynamics are characterized by three dimensionless parameters: the \peclet number $\Pe=\frac{\lambda}{\sqrt{R_d\gamma}}$, the carrying capacity $Z=\frac{\alpha-\beta}{\alpha}$, and the dimensionless reaction strength $s=\frac{Z\alpha}{\gamma}$. Here $\Pe$ quantifies activity as the ratio of the persistence length of active motion to the diffusion length over one tumbling time $\gamma^{-1}$, while $s$ compares the rate of birth/death events to tumbling. 

{\it Hydrodynamic Equations:} 
To derive the hydrodynamic evolution of allele frequencies in the one-dimensional active stepping-stone model, we use a standard path-integral approach~\cite{lefevre2007dynamics}. Starting from a system discretized in space and time and then taking the continuum limit, we provide full details in the Supplementary Material (SM~\ref{sec_fh_derivation}~\cite{SI}). Let $f_i^{\pm}(\tau_j) = n_i^{A^{\pm}}(\tau_j)/N$ be the local fraction of $A^{\pm}$ alleles at site $i$ and time $\tau_j$, where $n_i^{A^{\pm}}(\tau_j)$ and $n_i^B(\tau_j)$ are the numbers of $A^{\pm}$ and $B$ alleles, respectively. The total size of the local population is strictly conserved, $n_i^A(\tau_j) + n_i^B(\tau_j) = N$ for all $i,j$. The total fraction of active alleles is $f = f^+ + f^-$, so the passive fraction is $1 - f$. The local magnetization is $m_i(\tau_j) = f^+_i(\tau_j) - f^-_i(\tau_j)$. Generalizing to the two-dimensional square lattice, the particles are biased to move preferentially along their local polarization direction represented by four discrete states corresponding to the directions $+x$, $-x$, $+y$, and $-y$~\cite{jose2022active,jose2022first}. The magnetization is now a vector field $\mathbf{m} = (m_x,m_y)$  (see SM~\ref{sec:ASS_2d}~\cite{SI} for further details).

\begin{figure*}[t!]
    \centering  \includegraphics[width=1\textwidth]{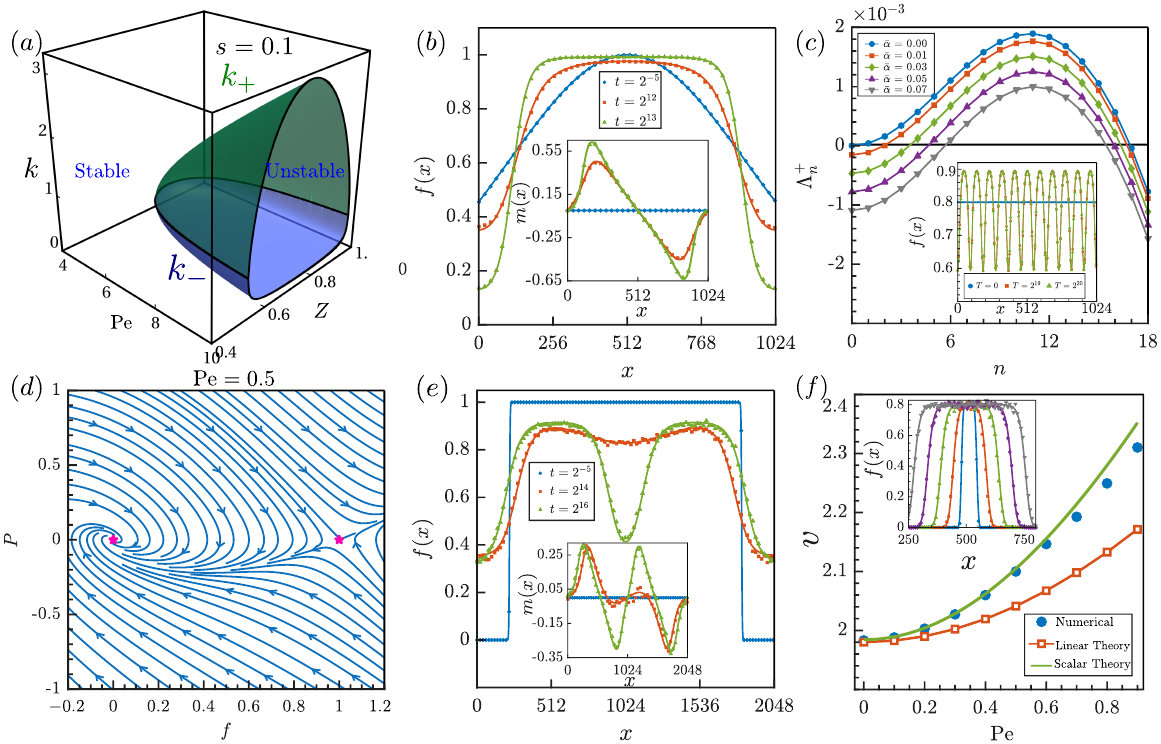}
\caption{Numerical verification of the hydrodynamic theory, Eq.~\eqref{eq:hydro2}. Distinct dynamical regimes are identified: $(a)$ A band of unstable modes is shown in the parameter space of $\Pe$ and $Z$ with $k_-$ and $k_+$ shown in different colors keeping $s=0.1$; $(b)$ activity-driven motility-induced phase separation (MIPS), characterized by macroscopic phase separation and spontaneous magnetization; $(c)$ a finite-$k$ instability characterized by an unstable band of Fourier modes, leading to a microphase-separated state, shown in the inset; $(d)$ phase portrait obtained from Eq.~\eqref{eq:position_momentum} at $\Pe=0.5$, showing the stable fixed point $(0,0)$ and the unstable fixed point $(1,0)$. $(e)$ a weak-reaction regime exhibiting finite-sized clusters and inhomogeneous magnetization; and (f) the wave regime, comparing the asymptotic velocity obtained from Eq.~\eqref{eq:hydro2} with the pulled-wave prediction, together with a representative traveling-wave profile in the inset. Parameters: $(a)$ $L=100$; $(b)$ $L=1024$, $\alpha=\beta=0$, $\lambda=50/L$, $\gamma=10/L^2$; $(c)$ $Z=0.8$, $\lambda=0.5$, $\gamma=0.01$, $L=1024$; $(d)$ $Z=1$, $s=1$; $(e)$ $Z=0.78$, $\lambda=0.02$, $\gamma=2.5\times10^{-5}$, $\alpha=2.5\times10^{-5}$; and $(f)$ $Z=0.8$, $\lambda=0.07$, $\gamma=0.01$, $\alpha=0.04$. The insets show the magnetization profiles in $(b)$ and $(e)$, the micro-phase-separated state in $(c)$ for $Z=0.8$, $\gamma=0.01$, and $\lambda=0.5$, $\bar{\alpha}=0.07$, and the traveling-wave profile at time instants $t=0,200, 400, 600, 800$ in $(f)$.
} 
\label{fig:profile}
\end{figure*}

During a microscopic time step $\Delta\tau$, the number of active particles at site $i$ changes by $\Delta n_i^A = 0, \pm 1$. Coarse-graining these dynamics under a local-equilibrium assumption~\cite{lefevre2007dynamics,bhutia2026diffusive} yields the macroscopic action and fluctuating hydrodynamic equations. To study universal behavior, we rescale space and time by the diffusion length and tumbling time, $x \rightarrow \sqrt{D_d/\gamma}\,x$ and $t \rightarrow t/\gamma$, obtaining dimensionless hydrodynamic equations. Their deterministic (noiseless) limit  is
\begin{eqnarray}
\partial_t f &= &\nabla^2 f
-\Pe\nabla\!\cdot\!\left[\mathbf{m}(1-f)\right]
+sf(1-f/Z), \nonumber\\
\partial_t\mathbf{m} &=& \nabla^2\mathbf{m}
-\Pe\nabla\!\left[f(1-f)\right]
-2\mathbf{m}+s(1-f/Z)\mathbf{m}.
~~~~~\label{eq:hydro2}
\end{eqnarray}
The explicit mapping between the transport coefficients $(\Pe, s)$ and the microscopic reaction-diffusion rates is given in SM~\ref{sec_fh_derivation}~\cite{SI}. Eq.~\eqref{eq:hydro2} provides the hydrodynamic description of the active stepping stone model in arbitrary dimensions. In the diffusion-dominated scaling regime, local configurations rapidly relax to a product Poisson measure~\cite{erignoux2016hydrodynamic,prakash2025exact,mukherjee2025hydrodynamics,bhutia2026diffusive}, where this hydrodynamic limit becomes exact. In particular, in the zero-reaction limit the ASM model reduces to that of Ref.~\cite{kourbane2018exact}, whose hydrodynamic limit has been proven to be exact.


{\it Verification of hydrodynamics in one dimension:} We first consider the one-dimensional case, which provides a simple
setting to test the predictions of the hydrodynamic theory against
microscopic simulations.
The hydrodynamic equation Eq.~\eqref{eq:hydro2} in one-dimension is:
\begin{eqnarray}
\partial_t \!f&=& \!\partial_x^2 f\!-\!\text{Pe} \partial_x \!\left[ m (1-f)\right]\!+ \!sf\!\left(1-\frac{f}{Z}\right),\nonumber\\
\partial_t m\!&=&\!\partial_x^2 m \!-\!\text{Pe} \partial_x \!\left[f (1-f)\right]\!-\!2m\!+\!s m\!\left(1-\frac{f}{Z}\right).
\label{eq:ALG_Fisher_Hydro_dimension_less}
\end{eqnarray}
In Fig.~\ref{fig:profile}, we present the solutions of these equations, which demonstrate that the microscopic simulations agree remarkably well with the predictions of the hydrodynamic theory described above. Macroscopic behavior is controlled by the three dimensionless parameters: $\Pe$, $s$, and $Z$. By varying $\Pe$, $s$, and $Z$ we obtain a range of steady states [Figs.~\ref{fig:profile}($(b),(c),(e),(f)$)]. In the $s=0$ limit, i.e., birth and death rates being equal, we obtain a phase diagram similar to Ref.~\cite{kourbane2018exact}. Low activity yields a spatially homogeneous, disordered phase with $m=0$ (outside the spinodal boundary), whereas high activity produces MIPS~\cite{cates2015motility,kourbane2018exact}: a dense domain accompanied by a strongly non-uniform magnetization profile [Fig.~\ref{fig:profile}($b$)]. With anisotropic birth and death ($s\neq 0$), three additional regimes appear. For strong reaction rates, demographic dynamics dominate, and the system relaxes to a homogeneous state with density uniformly approaching the carrying capacity $Z$ and vanishing magnetization. For weak reactions relative to activity, the MIPS state remains stable. At intermediate reaction strengths, however, the macroscopic domain becomes morphologically unstable and breaks into a steady state of smaller, localized clusters that locally retain MIPS-like features, i.e., a micro-phase-separated state [Fig.~\ref{fig:profile}($e$)]. In these clusters, the density can exceed $Z$; their stability reflects a dynamic balance in which reproduction dominates in low-density regions, and death prevails in high-density regions, while activity continuously transports newborn particles into high-density regions, sustaining the structures. Finally, when reaction and diffusion rates are comparable under low activity, the system supports a stable, propagating traveling-wave profile [Fig.~\ref{fig:profile}($f$)]. We note that the hydrodynamic description breaks down when active bias and diffusion become comparable. A detailed discussion of this breakdown is provided in the Supplementary Material (SM~\ref{breakdown_hydro}~\cite{SI}). 

Both compressible Toner-Tu and Malthusian flock models exhibit density inhomogeneities arising from the interplay of self-propelled convection and spin-spin alignment interactions~\cite{di2022evidence,toner2012birth}. In contrast, the active stepping stone model shows that a macroscopic phase separation can emerge without alignment, driven by the competition between activity (which amplifies density fluctuations) and proliferation (which suppresses them by restoring the density towards the carrying capacity).

{\it Linear stability analysis:}
We next investigate whether phase separation can occur starting from a homogeneous state. To this end, we expand Eq.~\eqref{eq:hydro2} around the steady state $f = Z$ and $m = 0$. We introduce small perturbations of the form $f = Z + \delta f$ and $m = \delta m$. Substituting these into Eq.~\eqref{eq:hydro2} and retaining only terms linear in the perturbations, we obtain the linearized equations.
We now assume monochromatic perturbations of the form $\delta f = \delta \hat{f}_k e^{i k x}$ and $\delta m = \delta \hat{m}_k e^{i k x}$. Defining the state vector $\vec{X} = (\delta \hat{f}_k, \delta \hat{m}_k)^{T}$, the evolution equation can be written compactly as $\partial_t \vec{X} = M \vec{X}$, where the stability matrix $M$ is given by
\begin{equation}
M =
\begin{pmatrix}
- k^2 -s & - i k\, \Pe\,(1 - Z) \\
- i k\, \Pe\,(1 - 2 Z) & -2 - k^2
\end{pmatrix}.
\end{equation}
The two eigenvalues are:
\begin{eqnarray}
\Lambda_k^{\pm}=\frac{1}{2}\left(-2 k^2 -s -2\pm\sqrt{\Delta}\right),
\label{eq:lam_pm}
\end{eqnarray}
where $\Delta=4 k^2 \Pe^2(2 Z-1)(1-Z)+(s -2)^2$.
The instability criterion can be written as $\Lambda_k^{\pm}>0$, which gives
\begin{equation}
\Pe^2 (1-Z)(2Z - 1) > 2+k^2+ s+\frac{2 s}{k^2}.
\label{eq:LS_1st_mode}
\end{equation}
The unstable band is given by $k_- < k < k_+$, where $k_\pm
=
\sqrt{\frac{A-(s+2)\pm \sqrt{\left(A-(s+2)\right)^2-8s}
}{2}
}$ and the constant A is defined as $A \equiv \Pe^{2}(2Z-1)(1-Z)$. Fig.~\ref{fig:profile}($a$) shows the band of unstable modes, with $k_-$ and $k_+$ shown in different colors. As $s$ increases, this unstable band narrows, with its boundaries shifting toward larger $\peclet$ values (see SM~\ref{Finite system size}~\cite{SI}). At the steady state, the maximum unstable mode is chosen for a larger reaction rate because the maximum eigenvalue just touches the y-axis (see Fig.~\ref{fig:profile}($c$)). As shown in Fig.~\ref{fig:profile}($c$) inset, the system evolves to a steady state in which only the $11^{\mathrm{th}}$ Fourier mode survives. This is consistent with the linear stability analysis shown in Fig.~\ref{fig:profile}($c$), where the maximum growth rate occurs around $n=11$. In contrast, for smaller reaction rates, all Fourier modes 
within the unstable band (see Fig.~\ref{fig:profile}($c$)) become unstable and interact through non-linear mode coupling. Consequently, the characteristic length scale selected in the steady state does not necessarily correspond to the fastest-growing linear mode (see~SM~\ref{kmax_1d}~\cite{SI}).

{\it Traveling waves in the presence of activity:}
In the absence of activity, Eq.~\eqref{eq:hydro2} reduces to the well-known FKPP equation, which admits traveling-wave solutions for sufficiently localized initial conditions~\cite{kolmogorov2019studies,brunet1997shift,doering2003interacting}. Weak activity preserves the qualitative structure of these waves but increases the propagation speed. At sufficiently large activity, or within the unstable regime shown in Fig.~\ref{fig:profile}($a$), the wave profile becomes strongly distorted and deviates from FKPP behavior. We determine the front velocity directly from the linearized hydrodynamic equations Eq.~\eqref{eq:hydro2} assuming a pulled wave approximation~\cite{kolmogorov2019studies,brunet1997shift}. The velocity is selected from the leading-edge dispersion relation.
Assuming that the wave has an exponentially decaying front of type $e^{-l (x - v t)}$, the marginal stability criterion~\cite{van1988front} selects the asymptotic velocity $v_{\min}=\min_l v(l)$, where the minimum lies along the curve $s=(l^3+\Pe~l^2+{\rm Pe})/(l+\Pe)$.
In the passive limit ($\Pe=0$), this reduces to the well-known FKPP result $v(l)=(l^2+s)/l$ and $v_{\min}=2\sqrt{D~s}$~\cite{SI}. The velocity of the front is also calculated by solving the full non-linear equation Eq.~\eqref{eq:hydro2} numerically. The linear theory shows a significant departure [Fig.~\ref{fig:profile}($f$)].

To understand the activity-induced speed enhancement, we introduce a dominant balance approximation, under which the polarization field is slaved to the density as $m(x,t)=-\frac{\lambda}{2\gamma}
\partial_x\left[f(1-f)\right]$
and Eq.~\eqref{eq:hydro2} reduces to a scalar field theory,
\begin{eqnarray}
\partial_t f
= \nabla\cdot(G[f]\nabla f)+s f(1-f/Z),
\label{eq:scalar_field}
\end{eqnarray}
where $G[f]=1+\kappa(1-f)(1-2f)$ and $\kappa=\Pe^2/2$. This approximation is valid for traveling-wave solutions (see SM~\ref{quasi-static}~\cite{SI}). 
Assuming a traveling-wave solution of the form $f(x,t)=f(\phi)$, with
$\phi=x-vt$, the partial differential equation
Eq.~\eqref{eq:scalar_field} reduces to an ordinary differential equation,
which can be recast as the dynamical system
\begin{eqnarray}
\dot{f}&=&P,\nonumber\\
\dot{P}&=&
\frac{
-vP
-\kappa(4f-3)P^2
-s f(1-f/Z)
}
{1+\kappa(1-f)(1-2f)},
\label{eq:position_momentum}
\end{eqnarray}
where the overdot denotes differentiation with respect to the
traveling-wave coordinate $\phi$, and $P=df/d\phi$ is the
conjugate momentum. For $Z=1$, the resulting phase portrait, shown in
Fig.~\ref{fig:profile}($d$), exhibits a heteroclinic orbit connecting the
fixed points $(0,0)$ and $(1,0)$ for a range of \peclet numbers. With increasing
activity, this connection is progressively lost, indicating a
modification of the selected wave speed. 

Using a pulled-wave ansatz for Eq.~\eqref{eq:scalar_field} and neglecting non-linearities at the leading edge~\cite{brunet2015exactly,doering2003interacting,brunet1997shift}, we obtain the asymptotic velocity $v=2\sqrt{D_{\rm eff}s}$ where $D_{\rm eff}=1+\kappa$. Fig.~\ref{fig:profile}($f$) compares the predictions of the scalar field theory Eq.~\eqref{eq:scalar_field} and the linear theory with numerical results of Eq.~\eqref{eq:hydro2}. The scalar theory captures the speed enhancement over a wide range of activity, while the linear theory increasingly deviates at large \peclet numbers. The eventual breakdown of the scalar theory at stronger activity can be understood from a phase-portrait analysis of the corresponding dynamical system. The denominator of Eq.~\eqref{eq:position_momentum} can vanish at finite
density, corresponding to the spinodal instability condition obtained
when the largest eigenvalue of the homogeneous zeroth mode becomes
positive. For example at $\Pe=10$, Eq.~\eqref{eq:position_momentum} predicts additional singular densities,
$f\simeq0.52$ and $f\simeq0.97$, appear in the physical interval,
where the denominator vanishes, and the phase trajectory becomes
discontinuous (see SM~\ref{active_wave}~\cite{SI}). This breakdown of the smooth heteroclinic orbit
indicates that the traveling-wave structure is strongly distorted at
large activity.  Although enhanced propagation speeds have also been observed in pushed waves, where non-linear growth in the bulk determines the velocity~\cite{birzu2018fluctuations,gandhi2016range}, the mechanism responsible for the speed enhancement in ASM is distinct and arises from activity-induced modifications of the effective transport. 

\begin{figure}[t!]
    \centering
    \includegraphics[width=1\linewidth]{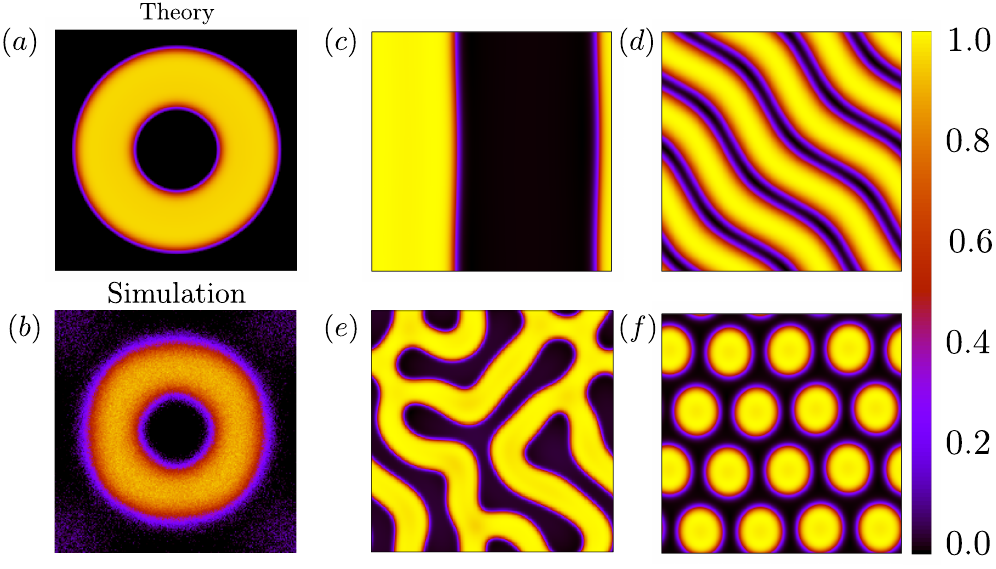}
    \caption{$(a),(b)$ Evolution of the two-dimensional density field at $t=2^{11}$ starting from a non-uniform initial state, comparing hydrodynamic solutions (a) with particle-level Monte Carlo simulations (b). Parameters: $L=200$, $Z=0.7$, $R_d=1$, $\alpha=0.001$, $\gamma=0.01$, and $\lambda=0.8$. $(c)$--$(f)$ MIPS, banded, ring-like, and hexagonal patterns obtained from numerical solution of Eq.~\eqref{eq:hydro2}. Here $L=256$, with (c) $Z=0.75$, $\Pe=7$, $\bar{\alpha}=0.01$; (d) $Z=0.75$, $\Pe=7$, $\bar{\alpha}=1$; (e) $Z=0.8$, $\Pe=7$, $\bar{\alpha}=0.1$ with a non uniform initial condition; and (f) $Z=0.7$, $\Pe=9$, $\bar{\alpha}=1$. Here $\bar{\alpha}=\alpha/\gamma$ and $\gamma=0.01$ for all the figures.}   
   \label{fig:2d_sim_th_plots}
\end{figure}


{\it Pattern formation in two dimensions:}
The stability of the homogeneous state is determined by the dispersion
relation $\Lambda^{\pm}(k)$. The state is stable when
$    \max_k \mathrm{Re}[\Lambda^{\pm}(k)]<0.$
A long-wavelength instability occurs when the unstable modes extend to
$k=0$. This
corresponds to the MIPS regime, where instability of the order of system size is
selected, and the domains can continue to coarsen~[Fig.~\ref{fig:2d_sim_th_plots}($c$)] up to the system size.
In contrast, a finite-wavelength instability occurs when
$\mathrm{Re}[\Lambda^{\pm}(k)]>0, k_-<k<k_+,k_m>0,$
where $k_m$ is the fastest-growing mode. This selects a characteristic
length scale $\ell_m=2\pi/k_m$ and can arrest coarsening. In two
dimensions, for an isotropic dispersion relation, the unstable modes
form an annulus in Fourier space,
$ k_-^2<k_x^2+k_y^2<k_+^2,$
with width $\Delta k=k_+-k_-$. The subsequent non-linear competition
between these modes determines the morphology, which may take the form
of bands, labyrinths, or hexagonal structures [see Figs.~\ref{fig:2d_sim_th_plots}($d$)-($f$)]. 

The full temporal evolution of the density profile from particle-level simulations matches remarkably well with the hydrodynamic theory (SM~\ref{sec:ASS_2d}~\cite{SI}). An initially localized circular density profile evolves into a distinct diamond-shaped structure by $t=2^{8}$, reflecting the fourfold symmetry of the underlying lattice and the dominance of active-current directionality. At later times, intense localized competition and net demographic death within the dense core lead to a density depletion, forming a central hole. The successful prediction of both the diamond-shaped morphology and this hollow-core transition demonstrates that our hydrodynamic theory quantitatively captures the non-equilibrium coupling between motility and proliferation. 

Figs.~\ref{fig:2d_sim_th_plots}($a$),($b$) are the two-dimensional analogs of Fig.~\ref{fig:profile}($e$), where a large dense cluster breaks due to the net death rate within the high-density region. The overall mean-field equation predicts similar behavior in both dimensions. This is because the chosen parameters lie deep inside the spinodal region, where the deterministic hydrodynamic instability dominates over noise, leading to robust phase separation.
Furthermore, this 2D framework predicts the emergence of micro-phase separation~[Fig.~\ref{fig:2d_sim_th_plots}($f$)] with a hexagonal lattice structure, a phenomenon observed in chemotactic bacterial colonies both in experiments~\cite{budrene1991complex,budrene1995dynamics} and also in other theoretical models~\cite{hernandez2004clustering,silvano2026hydrodynamic,takamura2025spatially}. 
The model also shows bulk phase separation and a band-like structure, labyrinth-like morphology, and a finite number of clusters arranged in a periodic pattern, as shown in Figs.~\ref{fig:2d_sim_th_plots}($c$)-($f$) at steady state. In the steady state, the relation $\langle f\rangle=\langle f^2\rangle/Z$ holds, where $\langle \rangle$ denotes the spatial average (see SM~\ref{steady state}~\cite{SI} for details). Depending on the initial conditions and parameter regimes, the model also supports a rich family of spatial patterns (detailed in SM~\ref{sec:ASS_2d}~\cite{SI}). We note that ring-like and hexagonal patterns in proliferating active matter have been reported previously in Refs.~\cite{curatolo2020cooperative,cates2010arrested,takamura2025spatially,hernandez2004clustering}. Here we obtain these morphologies from a distinct fixed-occupancy stepping-stone model with active lattice-gas migration, and derive the corresponding hydrodynamic description from the microscopic dynamics.

Finally, to obtain a system-size-independent description, diffusive rescaling can be performed, with the spatial and temporal coordinates transformed as
$x\rightarrow x/L$ and $t\rightarrow t/L^2$. Accordingly, the microscopic reaction parameters are rescaled as
$\lambda\rightarrow\lambda/L$, $\gamma\rightarrow\gamma/L^2$, and
$\alpha\rightarrow\alpha/L^2,\beta \rightarrow \beta/L^2$. Under this scaling, the reaction
parameters are small compared to diffusion in the
large-$L$ limit, consistent with the regime in which hydrodynamic
limits are known to be exact~\cite{erignoux2016hydrodynamic,prakash2025exact,mukherjee2025hydrodynamics,mason2023macroscopic}.

{\it Discussion:} 
In this Letter, we have introduced a microscopic model that couples active motility with demographic birth-death dynamics in a proliferating population. Starting from microscopic rules, we derived the corresponding non-linear hydrodynamic equations and validated them via Monte Carlo simulations and numerical solutions. 
This model exhibits disordered, phase-separated, wave-propagating, and microphase-separated states. Weak reactions yield MIPS, whereas comparable reaction and diffusion timescales produce robust propagating waves. At intermediate reaction strengths, stable microphase-separated clusters form, resembling patterns observed in chemotactic bacteria~\cite{budrene1991complex,budrene1995dynamics}. We also find FKPP-type invasion waves, where weak activity accelerates fronts, consistent with marginal stability analysis. However, strong activity destabilizes these fronts, which can naturally be understood using a phase portrait analysis.

Our framework provides a microscopic foundation for the coupled transport and demographic dynamics underlying active growth, placing conserved active transport and non-conserved birth–death fluctuations within recent symmetry- and conservation-law-based extensions of the Hohenberg–Halperin classification of active matter \cite{hohenberg1977theory,cates2025active}. In this sense, the model provides a microscopic realization at the intersection of active Model AB \cite{li2021hierarchical,mondal2026derivation} and Malthusian flocking dynamics \cite{toner2012birth}, while extending the same coarse-graining framework to non-reciprocal interacting fields \cite{saha2020scalar,brauns2024nonreciprocal}. The non-linear hydrodynamic equations derived in this work therefore offer a route to uncovering broader universality classes governing non-equilibrium spatial evolution.

{\it Acknowledgments:} We thank S.~Jose, S.~Prakash, S.~Saryal, V.~Subashri, P.~Chaudhuri, A.~Maitra and H.~L\"owen for discussions. This project was funded by intramural funds at TIFR Hyderabad from the Department of Atomic Energy (DAE), Government of India, under Project Identification No. RTI4007.
\bibliography{Reference_v2}

@article{marchetti2013hydrodynamics,
  title={Hydrodynamics of soft active matter},
  author={Marchetti, M. Cristina and Joanny, Jean-Fran{\c{c}}ois and Ramaswamy, Sriram and Liverpool, Tanniemola B. and Prost, Jacques and Rao, Madan and Simha, R. Aditi},
  journal={\href{https://journals.aps.org/rmp/abstract/10.1103/RevModPhys.85.1143}{Rev. Mod. Phys.}},
  volume={85},
  number={3},
  pages={1143--1189},
  year={2013},
  publisher={APS}
}

@article{te2026colloquium,
  title={Colloquium: What do we mean by ‘active matter’?},
  author={te Vrugt, Michael and Liebchen, Benno and Cates, Michael E},
  journal={\href{https://journals.aps.org/rmp/abstract/10.1103/wd4f-q7kv}{Rev. Mod. Phys.}},
  volume={98},
  number={3},
  pages={031001},
  year={2026},
  publisher={APS}
}

@article{liu2019self,
  title={Self-driven phase transitions drive Myxococcus xanthus fruiting body formation},
  author={Liu, Guannan and Patch, Adam and Bahar, Fatmag{\"u}l and Yllanes, David and Welch, Roy D and Marchetti, M Cristina and Thutupalli, Shashi and Shaevitz, Joshua W},
  journal={\href{https://journals.aps.org/prl/abstract/10.1103/PhysRevLett.122.248102}{Phys. Rev. Lett}},
  volume={122},
  number={24},
  pages={248102},
  year={2019},
  publisher={APS}
}

@misc{SI,
Note = {See Supplemental Material for details}}

@article{brunet2013genealogies,
  title={Genealogies in simple models of evolution},
  author={Brunet, {\'E}ric and Derrida, Bernard},
  journal={\href{https://iopscience.iop.org/article/10.1088/1742-5468/2013/01/P01006}{Journal of Statistical Mechanics: Theory and Experiment}},
  volume={2013},
  number={01},
  pages={P01006},
  year={2013},
  publisher={IOP Publishing and SISSA}
}

@article{brunet2008universal,
  title={Universal tree structures in directed polymers and models of evolving populations},
  author={Brunet, {\'E}ric and Derrida, Bernard and Simon, Damien},
  journal={\href{https://journals.aps.org/pre/abstract/10.1103/PhysRevE.78.061102}{Phys. Rev. E}},
  volume={78},
  number={6},
  pages={061102},
  year={2008},
  publisher={APS}
}

@article{sanchez2012spontaneous,
  title={Spontaneous motion in hierarchically assembled active matter},
  author={Sanchez, Tim and Chen, Daniel T. N. and DeCamp, Stephen J. and Heymann, Michael and Dogic, Zvonimir},
  journal={\href{https://www.nature.com/articles/nature11591}{Nature}},
  volume={491},
  number={7424},
  pages={431--434},
  year={2012},
  publisher={Nature Publishing Group UK London}
}

@article{narayan2007long,
  title={Long-lived giant number fluctuations in a swarming granular nematic},
  author={Narayan, Vijay and Ramaswamy, Sriram and Menon, Narayanan},
  journal={\href{https://www.science.org/doi/10.1126/science.1140414}{Science}},
  volume={317},
  number={5834},
  pages={105--108},
  year={2007},
  publisher={American Association for the Advancement of Science}
}

@article{toner1995long,
  title={Long-range order in a two-dimensional dynamical XY model: How birds fly together},
  author={Toner, John and Tu, Yuhai},
  journal={\href{https://journals.aps.org/prl/abstract/10.1103/PhysRevLett.75.4326}{Phys. Rev. Lett.}},
  volume={75},
  number={23},
  pages={4326--4329},
  year={1995},
  publisher={APS}
}

@article{cavagna2010scale,
  title={Scale-free correlations in starling flocks},
  author={Cavagna, Andrea and Cimarelli, Alessio and Giardina, Irene and Parisi, Giorgio and Santagati, Raffaele and Stefanini, Fabio and Viale, Massimiliano},
  journal={\href{https://www.pnas.org/doi/10.1073/pnas.1005766107}{Proc. Natl. Acad. Sci. U.S.A.}},
  volume={107},
  number={26},
  pages={11865--11870},
  year={2010},
  publisher={National Academy of Sciences}
}

@article{jain2024inertia,
  title={Inertia drives concentration-wave turbulence in swimmer suspensions},
  author={Jain, Purnima and Rana, Navdeep and Ramaswamy, Sriram and Perlekar, Prasad},
  journal={\href{https://journals.aps.org/prl/abstract/10.1103/PhysRevLett.133.158302}{Phys. Rev. Lett.}},
  volume={133},
  number={15},
  pages={158302},
  year={2024},
  publisher={APS}
}

@article{ramola2014universal,
  title={Universal order and gap statistics of critical branching Brownian motion},
  author={Ramola, Kabir and Majumdar, Satya N. and Schehr, Gr{\'e}gory},
  journal={\href{https://journals.aps.org/prl/abstract/10.1103/PhysRevLett.112.210602}{Phys. Rev. Lett.}},
  volume={112},
  number={21},
  pages={210602},
  year={2014},
  publisher={APS}
}

@article{ramola2015spatial,
  title={Spatial extent of branching Brownian motion},
  author={Ramola, Kabir and Majumdar, Satya N. and Schehr, Gr{\'e}gory},
  journal={\href{https://journals.aps.org/pre/abstract/10.1103/PhysRevE.91.042131}{Phys. Rev. E}},
  volume={91},
  number={4},
  pages={042131},
  year={2015},
  publisher={APS}
}

@article{hallatschek2023proliferating,
  title={Proliferating active matter},
  author={Hallatschek, Oskar and Datta, Sujit S. and Drescher, Knut and Dunkel, J{\"o}rn and Elgeti, Jens and Waclaw, Bartek and Wingreen, Ned S.},
  journal={\href{https://www.nature.com/articles/s42254-023-00593-0}{Nat. Rev. Phys.}},
  volume={5},
  number={7},
  pages={407--419},
  year={2023},
  publisher={Nature Publishing Group UK London}
}

@article{korolev2010genetic,
  title={Genetic demixing and evolution in linear stepping stone models},
  author={Korolev, Kirill S. and Avlund, Mikkel and Hallatschek, Oskar and Nelson, David R.},
  journal={\href{https://journals.aps.org/rmp/abstract/10.1103/RevModPhys.82.1691}{Rev. Mod. Phys.}},
  volume={82},
  number={2},
  pages={1691--1718},
  year={2010},
  publisher={APS}
}

@article{cates2025active,
  title={Active phase separation: New phenomenology from non-equilibrium physics},
  author={Cates, Michael E. and Nardini, Cesare},
  journal={\href{https://iopscience.iop.org/article/10.1088/1361-6633/add278}{Rep. Prog. Phys.}},
  volume={88},
  number={5},
  pages={056601},
  year={2025},
  publisher={IOP Publishing}
}

@article{toner2012birth,
  title={Birth, death, and flight: A theory of Malthusian flocks},
  author={Toner, John},
  journal={\href{https://journals.aps.org/prl/abstract/10.1103/PhysRevLett.108.088102}{Phys. Rev. Lett.}},
  volume={108},
  number={8},
  pages={088102},
  year={2012},
  publisher={APS}
}

@article{yadav2025coarsening,
  title={Coarsening kinetics in active model B+: Macroscale and microscale phase separation},
  author={Yadav, Pradeep Kumar and Mishra, Shradha and Puri, Sanjay},
  journal={\href{https://journals.aps.org/pre/abstract/10.1103/kb9k-w7jr}{Phys. Rev. E}},
  volume={112},
  number={3},
  pages={035412},
  year={2025},
  publisher={APS}
}

@article{cates2015motility,
  title={Motility-induced phase separation},
  author={Cates, Michael E. and Tailleur, Julien},
  journal={\href{https://www.annualreviews.org/content/journals/10.1146/annurev-conmatphys-031214-014710}{Annu. Rev. Condens. Matter Phys.}},
  volume={6},
  number={1},
  pages={219--244},
  year={2015},
  publisher={Annual Reviews}
}

@article{curatolo2020cooperative,
  title={Cooperative pattern formation in multi-component bacterial systems through reciprocal motility regulation},
  author={Curatolo, A. I. and Zhou, Norman and Zhao, Yongfeng and Liu, Chenli and Daerr, Adrian and Tailleur, Julien and Huang, Jiandong},
  journal={\href{https://www.nature.com/articles/s41567-020-0964-z}{Nat. Phys.}},
  volume={16},
  number={11},
  pages={1152--1157},
  year={2020},
  publisher={Nature Publishing Group UK London}
}

@article{soares2019polarity,
  title={Polarity in ciliate models: From cilia to cell architecture},
  author={Soares, Helena and Carmona, Bruno and Nolasco, Sofia and Viseu Melo, Lu{\'\i}s},
  journal={\href{https://www.frontiersin.org/journals/cell-and-developmental-biology/articles/10.3389/fcell.2019.00240/full}{Front. Cell Dev. Biol.}},
  volume={7},
  pages={240},
  year={2019},
  publisher={Frontiers Media SA}
}

@article{wittkowski2014scalar,
  title={Scalar $\varphi^4$ field theory for active-particle phase separation},
  author={Wittkowski, Raphael and Tiribocchi, Adriano and Stenhammar, Joakim and Allen, Rosalind J. and Marenduzzo, Davide and Cates, Michael E.},
  journal={\href{https://www.nature.com/articles/ncomms5351}{Nat. Commun.}},
  volume={5},
  number={1},
  pages={4351},
  year={2014},
  publisher={Nature Publishing Group UK London}
}

@article{meissner2024introduction,
  title={An introduction to phase ordering in scalar active matter},
  author={Meissner, Laura and Yeomans, Julia M.},
  journal={\href{https://link.springer.com/article/10.1140/epjs/s11734-024-01273-5}{Eur. Phys. J. Spec. Top.}},
  volume={233},
  number={17},
  pages={2701--2710},
  year={2024},
  publisher={Springer}
}

@article{cates2010arrested,
  title={Arrested phase separation in reproducing bacteria creates a generic route to pattern formation},
  author={Cates, Michael E. and Marenduzzo, D. and Pagonabarraga, I. and Tailleur, J.},
  journal={\href{https://www.pnas.org/doi/10.1073/pnas.1001994107}{Proc. Natl. Acad. Sci. U.S.A.}},
  volume={107},
  number={26},
  pages={11715--11720},
  year={2010},
  publisher={National Academy of Sciences}
}

@article{budrene1991complex,
  title={Complex patterns formed by motile cells of Escherichia coli},
  author={Budrene, Elena O. and Berg, Howard C.},
  journal={\href{https://www.nature.com/articles/349630a0}{Nature}},
  volume={349},
  number={6310},
  pages={630--633},
  year={1991},
  publisher={Nature Publishing Group UK London}
}

@article{budrene1995dynamics,
  title={Dynamics of formation of symmetrical patterns by chemotactic bacteria},
  author={Budrene, Elena O. and Berg, Howard C.},
  journal={\href{https://www.nature.com/articles/376049a0}{Nature}},
  volume={376},
  number={6535},
  pages={49--53},
  year={1995},
  publisher={Nature Publishing Group UK London}
}

@article{ben1997snowflake,
  title={From snowflake formation to growth of bacterial colonies II: Cooperative formation of complex colonial patterns},
  author={Ben-Jacob, Eshel},
  journal={\href{https://www.tandfonline.com/doi/abs/10.1080/001075197182405}{Contemp. Phys.}},
  volume={38},
  number={3},
  pages={205--241},
  year={1997},
  publisher={Taylor \& Francis}
}

@article{kourbane2018exact,
  title={Exact hydrodynamic description of active lattice gases},
  author={Kourbane-Houssene, Mourtaza and Erignoux, Cl{\'e}ment and Bodineau, Thierry and Tailleur, Julien},
  journal={\href{https://journals.aps.org/prl/pdf/10.1103/PhysRevLett.120.268003}{Phys. Rev. Lett.}},
  volume={120},
  number={26},
  pages={268003},
  year={2018},
  publisher={APS}
}

@article{pigolotti2013growth,
  title={Growth, competition and cooperation in spatial population genetics},
  author={Pigolotti, Simone and Benzi, Roberto and Perlekar, Prasad and Jensen, Mogens H{\o}gh and Toschi, Federico and Nelson, David R.},
  journal={\href{https://www.sciencedirect.com/science/article/abs/pii/S0040580912001360}{Theor. Popul. Biol.}},
  volume={84},
  pages={72--86},
  year={2013},
  publisher={Elsevier}
}

@article{singha2020fixation,
  title={Fixation in competing populations: Diffusion and strategies for survival},
  author={Singha, Tapas and Perlekar, Prasad and Barma, Mustansir},
  journal={\href{https://journals.aps.org/prresearch/pdf/10.1103/PhysRevResearch.2.023412}{Phys. Rev. Res.}},
  volume={2},
  number={2},
  pages={023412},
  year={2020},
  publisher={APS}
}

@book{Moran,
  author={Moran, P. A. P.},
  title={\href{https://www.cambridge.org/core/journals/mathematical-proceedings-of-the-cambridge-philosophical-society/article/abs/random-processes-in-genetics/9EEED52D6AE22A026036F32D9B1CA07C}{Random Processes in Genetics}},
  year={1958},
  publisher={Cambridge University Press}
}

@article{kimura1964stepping,
  title={The stepping stone model of population structure and the decrease of genetic correlation with distance},
  author={Kimura, Motoo and Weiss, George H.},
  journal={\href{https://www.ncbi.nlm.nih.gov/pmc/articles/PMC1210594/pdf/561.pdf}{Genetics}},
  volume={49},
  number={4},
  pages={561--576},
  year={1964},
  publisher={Oxford University Press}
}

@article{lefevre2007dynamics,
  title={Dynamics of interacting particle systems: Stochastic process and field theory},
  author={Lefevre, Alexandre and Biroli, Giulio},
  journal={\href{https://iopscience.iop.org/article/10.1088/1742-5468/2007/07/P07024}{J. Stat. Mech.}},
  volume={2007},
  number={07},
  pages={P07024},
  year={2007},
  publisher={IOP Publishing}
}

@article{li2021hierarchical,
  title={Hierarchical microphase separation in non-conserved active mixtures},
  author={Li, Yuting I. and Cates, Michael E.},
  journal={\href{https://link.springer.com/article/10.1140/epje/s10189-021-00113-x}{Eur. Phys. J. E}},
  volume={44},
  number={9},
  pages={119},
  year={2021},
  publisher={Springer}
}

@incollection{kolmogorov2019studies,
  title={Studies of the diffusion with the increasing quantity of the substance; its application to a biological problem},
  author={Kolmogorov, Andrey Nikolaevich and Piskunov, N. S.},
  booktitle={\href{https://benjaminmoll.com/wp-content/uploads/2019/07/KPP.pdf}{Differential Equations}},
  pages={106--132},
  year={2019},
  publisher={CRC Press}
}

@article{martin1973statistical,
  title={Statistical dynamics of classical systems},
  author={Martin, Paul Cecil and Siggia, E. D. and Rose, H. A.},
  journal={\href{https://journals.aps.org/pra/abstract/10.1103/PhysRevA.8.423}{Phys. Rev. A}},
  volume={8},
  number={1},
  pages={423--437},
  year={1973},
  publisher={APS}
}

@article{van1988front,
  title={Front propagation into unstable states: Marginal stability as a dynamical mechanism for velocity selection},
  author={van Saarloos, Wim},
  journal={\href{https://journals.aps.org/pra/abstract/10.1103/PhysRevA.37.211}{Phys. Rev. A}},
  volume={37},
  number={1},
  pages={211--229},
  year={1988},
  publisher={APS}
}

@article{brunet1997shift,
  title={Shift in the velocity of a front due to a cutoff},
  author={Brunet, Eric and Derrida, Bernard},
  journal={\href{https://journals.aps.org/pre/pdf/10.1103/PhysRevE.56.2597}{Phys. Rev. E}},
  volume={56},
  number={3},
  pages={2597--2604},
  year={1997},
  publisher={APS}
}

@article{brunet2015exactly,
  title={An exactly solvable travelling wave equation in the Fisher--KPP class},
  author={Brunet, {\'E}ric and Derrida, Bernard},
  journal={J. Stat. Phys.},
  volume={161},
  number={4},
  pages={801--820},
  year={2015},
  publisher={Springer}
}

@article{jose2022active,
  title={Active random walks in one and two dimensions},
  author={Jose, Stephy and Mandal, Dipanjan and Barma, Mustansir and Ramola, Kabir},
  journal={\href{https://journals.aps.org/pre/abstract/10.1103/PhysRevE.105.064103}{Phys. Rev. E}},
  volume={105},
  number={6},
  pages={064103},
  year={2022},
  publisher={APS}
}

@article{hallatschek2010life,
  title={Life at the front of an expanding population},
  author={Hallatschek, Oskar and Nelson, David R.},
  journal={\href{https://pubmed.ncbi.nlm.nih.gov/19682067}{Evolution}},
  volume={64},
  number={1},
  pages={193--206},
  year={2010},
  publisher={Blackwell Publishing}
}

@article{jose2023current,
  title={Current fluctuations in an interacting active lattice gas},
  author={Jose, Stephy and Dandekar, Rahul and Ramola, Kabir},
  journal={\href{https://doi.org/10.1103/PhysRevE.110.064153}{J. Stat. Mech.}},
  volume={2023},
  number={8},
  pages={083208},
  year={2023},
  publisher={IOP Publishing}
}

@article{doering2003interacting,
  title={Interacting particles, the stochastic Fisher--Kolmogorov--Petrovsky--Piscounov equation, and duality},
  author={Doering, Charles R. and Mueller, Carl and Smereka, Peter},
  journal={\href{https://people.math.rochester.edu/faculty/cmlr/Preprints/inter-fkpp.pdf}{Physica A}},
  volume={325},
  number={1--2},
  pages={243--259},
  year={2003},
  publisher={Elsevier}
}

@article{saha2023current,
  title={Current fluctuations in a semi-infinite line},
  author={Saha, Soumyabrata and Sadhu, Tridib},
  journal={\href{https://iopscience.iop.org/article/10.1088/1742-5468/ace3b2/meta}{J. Stat. Mech.}},
  volume={2023},
  number={7},
  pages={073207},
  year={2023},
  publisher={IOP Publishing}
}

@article{pechenik1999interfacial,
  title={Interfacial velocity corrections due to multiplicative noise},
  author={Pechenik, Leonid and Levine, Herbert},
  journal={\href{https://journals.aps.org/pre/pdf/10.1103/PhysRevE.59.3893}{Phys. Rev. E}},
  volume={59},
  number={4},
  pages={3893--3900},
  year={1999},
  publisher={APS}
}

@article{mukherjee2025hydrodynamics,
  title={Hydrodynamics of a hard-core active lattice gas},
  author={Mukherjee, Ritwik and Saha, Soumyabrata and Sadhu, Tridib and Dhar, Abhishek and Sabhapandit, Sanjib},
  journal={\href{https://journals.aps.org/pre/abstract/10.1103/PhysRevE.111.024128}{Phys. Rev. E}},
  volume={111},
  number={2},
  pages={024128},
  year={2025},
  publisher={APS}
}

@article{agranov2021exact,
  title={Exact fluctuating hydrodynamics of active lattice gases—typical fluctuations},
  author={Agranov, Tal and Ro, Sunghan and Kafri, Yariv and Lecomte, Vivien},
  journal={\href{https://iopscience.iop.org/article/10.1088/1742-5468/ac1406}{J. Stat. Mech. }},
  volume={2021},
  number={8},
  pages={083208},
  year={2021},
  publisher={IOP Publishing and SISSA}
}

@article{prakash2025exact,
  title={Exact fluctuating hydrodynamics of the scaled light--heavy model},
  author={Prakash, Shilpa and Barma, Mustansir and Ramola, Kabir},
  journal={\href{https://iopscience.iop.org/article/10.1088/1742-5468/ae09a3}{J. Stat. Mech.}},
  volume={2025},
  number={10},
  pages={103208},
  year={2025},
  publisher={IOP Publishing}
}

@article{mason2023macroscopic,
  author  = {Mason, James and Jack, Robert L. and Bruna, Maria},
  title   = {Macroscopic Behaviour in a Two-Species Exclusion Process via the Method of Matched Asymptotics},
  journal = {\href{https://link.springer.com/article/10.1007/s10955-022-03036-9}{J. Stat. Phys.}},
  volume  = {190},
  number  = {3},
  pages   = {47},
  year    = {2023}
}

@article{hohenberg1977theory,
  title={Theory of dynamic critical phenomena},
  author={Hohenberg, Pierre C. and Halperin, Bertrand I.},
  journal={\href{https://journals.aps.org/rmp/abstract/10.1103/RevModPhys.49.435}{Rev. Mod. Phys.}},
  volume={49},
  number={3},
  pages={435--479},
  year={1977},
  publisher={APS}
}

@article{erignoux2016hydrodynamic,
  title={Hydrodynamic limit for an active exclusion process},
  author={Erignoux, Cl{\'e}ment},
  journal={\href{https://arxiv.org/abs/1608.04937}{arXiv preprint arXiv:1608.04937}},
  year={2016}
}

@phdthesis{duty2000broken,
  title={Broken symmetry and critical phenomena in population genetics: The stepping-stone model},
  author={Duty, Timothy Lee},
  year={2000},
  school={\href{https://open.library.ubc.ca/media/stream/pdf/831/1.0085696/2}{University of British Columbia}}
}

@article{wright1931evolution,
  title={Evolution in Mendelian populations},
  author={Wright, Sewall},
  journal={\href{https://acteon.webs.upv.es/ARTICULOS/Wright.pdf}{Genetics}},
  volume={16},
  number={2},
  pages={97--159},
  year={1931}
}

@article{mondal2026derivation,
  title={Derivation and analysis of the amplitude equation for generalized active model B in the presence of chemical reactions},
  author={Mondal, Sayantan and Das, Prasenjit},
  journal={\href{https://journals.aps.org/pre/abstract/10.1103/s7sp-d3hm}{Phys. Rev. E}},
  volume={113},
  number={5},
  pages={054139},
  year={2026},
  publisher={APS}
}

@article{doi1976second,
  title={Second quantization representation for classical many-particle system},
  author={Doi, Masao},
  journal={\href{https://iopscience.iop.org/article/10.1088/0305-4470/9/9/008}{J. Phys. A}},
  volume={9},
  number={9},
  pages={1465--1477},
  year={1976},
  publisher={IOP Publishing}
}

@article{peliti1985path,
  title={Path integral approach to birth-death processes on a lattice},
  author={Peliti, Luca},
  journal={\href{http://www.peliti.org/Publications/pathIntegrals.pdf}{J. Phys.}},
  volume={46},
  number={9},
  pages={1469--1483},
  year={1985},
  publisher={Soci{\'e}t{\'e} Fran{\c{c}}aise de Physique}
}

@article{hallatschek2009fisher,
  title={Fisher waves in the strong noise limit},
  author={Hallatschek, Oskar and Korolev, Kirill S.},
  journal={\href{https://journals.aps.org/prl/pdf/10.1103/PhysRevLett.103.108103}{Phys. Rev. Lett.}},
  volume={103},
  number={10},
  pages={108103},
  year={2009},
  publisher={APS}
}

@phdthesis{martin2021nonequilibrium,
  title={Nonequilibrium signatures and phase transitions in active matter and beyond},
  author={Martin, David},
  year={2021},
  school={\href{https://theses.hal.science/tel-03479496/}{Universit{\'e} Paris Cit{\'e}}}
}

@book{cates2019active,
  title={Active field theories},
  author={Cates, Michael E.},
  year={2019},
  publisher={\href{https://arxiv.org/pdf/1904.01330}{arXiv:1904.01330}}
}

@article{thompson2011lattice,
  title={Lattice models of nonequilibrium bacterial dynamics},
  author={Thompson, Alasdair G. and Tailleur, Julien and Cates, Michael E. and Blythe, Richard A.},
  journal={\href{https://iopscience.iop.org/article/10.1088/1742-5468/2011/02/P02029/meta}{J. Stat. Mech.}},
  volume={2011},
  number={02},
  pages={P02029},
  year={2011},
  publisher={IOP Publishing}
}

@article{saha2020scalar,
  title={Scalar active mixtures: The nonreciprocal Cahn-Hilliard model},
  author={Saha, Suropriya and Agudo-Canalejo, Jaime and Golestanian, Ramin},
  journal={\href{https://journals.aps.org/prx/abstract/10.1103/PhysRevX.10.041009}{Phys. Rev. X}},
  volume={10},
  number={4},
  pages={041009},
  year={2020},
  publisher={APS}
}

@article{huang2024active,
  title={Active pattern formation emergent from single-species nonreciprocity},
  author={Huang, Zhi-Feng and Vrugt, Michael te and Martins, Jonas Mayer and Wittkowski, Raphael and L{\"o}wen, Hartmut},
  journal={arXiv preprint arXiv:2404.10093},
  year={2024}
}

@article{brauns2024nonreciprocal,
  title={Nonreciprocal pattern formation of conserved fields},
  author={Brauns, Fridtjof and Marchetti, M. Cristina},
  journal={\href{https://journals.aps.org/prx/abstract/10.1103/PhysRevX.14.021014}{Phys. Rev. X}},
  volume={14},
  number={2},
  pages={021014},
  year={2024},
  publisher={APS}
}

@article{jose2022first,
  title={First passage statistics of active random walks on one and two dimensional lattices},
  author={Jose, Stephy},
  journal={Journal of Statistical Mechanics: Theory and Experiment},
  volume={2022},
  number={11},
  pages={113208},
  year={2022},
  publisher={IOP Publishing and SISSA}
}

@article{di2022evidence,
  title={Evidence of fluctuation-induced first-order phase transition in active matter},
  author={Di Carlo, Luca and Scandolo, Mattia},
  journal={New J. Phys.},
  volume={24},
  number={12},
  pages={123032},
  year={2022},
  publisher={IOP Publishing}
}

@article{bhutia2026diffusive,
  title={Diffusive Noise Controls Early Stages of Genetic Demixing},
  author={Bhutia, Rashmiranjan and Jose, Stephy and Perlekar, Prasad and Ramola, Kabir},
  journal={\href{https://journals.aps.org/prl/abstract/10.1103/xmbp-d85n}{Phys. Rev. Lett}},
  volume={137},
  number={9},
  pages={098401},
  year={2026},
  publisher={American Physical Society (APS)},
}

@article{birzu2018fluctuations,
  title={Fluctuations uncover a distinct class of traveling waves},
  author={Birzu, Gabriel and Hallatschek, Oskar and Korolev, Kirill S.},
  journal={\href{https://www.pnas.org/doi/10.1073/pnas.1715737115}{Proc. Natl. Acad. Sci. U.S.A.}},
  volume={115},
  number={16},
  pages={E3645--E3654},
  year={2018},
  publisher={National Academy of Sciences}
}

@article{gandhi2016range,
  title={Range expansions transition from pulled to pushed waves as growth becomes more cooperative in an experimental microbial population},
  author={Gandhi, Saurabh R. and Yurtsev, Eugene Anatoly and Korolev, Kirill S. and Gore, Jeff},
  journal={\href{https://www.pnas.org/doi/10.1073/pnas.1521056113}{Proc. Natl. Acad. Sci. U.S.A.}},
  volume={113},
  number={25},
  pages={6922--6927},
  year={2016},
  publisher={National Academy of Sciences}
}

@article{silvano2026hydrodynamic,
  title={Hydrodynamic description of proliferating active matter},
  author={Silvano, Nathan and Hernandez-Garcia, Emilio and L{\'o}pez, Crist{\'o}bal},
  journal={\href{https://arxiv.org/abs/2608.03317}{arXiv preprint arXiv:2608.03317}},
  year={2026}
}

@article{hernandez2004clustering,
  title={Clustering, advection, and patterns in a model of population dynamics with neighborhood-dependent rates},
  author={Hern{\'a}ndez-Garc{\'\i}a, Emilio and L{\'o}pez, Crist{\'o}bal},
  journal={\href{https://journals.aps.org/pre/abstract/10.1103/PhysRevE.70.016216}{Phys. Rev. E}},
  volume={70},
  number={1},
  pages={016216},
  year={2004},
  publisher={APS}
}

@article{takamura2025spatially,
  title={Spatially structured flocking in a proliferating population of self-propelled organisms without explicit alignment interactions},
  author={Takamura, Seiya and Saito, Nen},
  journal={\href{https://journals.aps.org/prresearch/abstract/10.1103/v3gm-vhry}{Phys. Rev. Res.}},
  volume={7},
  number={3},
  pages={033299},
  year={2025},
  publisher={APS}
}

\clearpage

\begin{widetext}

\section*{\large Supplemental Material for\\ ``Traveling waves, phase separation and pattern formation in\\ the active stepping stone model''}
This document provides supplemental figures and details related to the results presented in the main text.

\maketitle


\section{Derivation of Fluctuating Hydrodynamics equations}
 \label{sec_fh_derivation}
 There are several approaches to derive fluctuating hydrodynamics starting from microscopic models~\cite{doi1976second,peliti1985path,lefevre2007dynamics,martin2021nonequilibrium}. Here, we follow the method introduced in Ref.~\cite{lefevre2007dynamics}, which has been applied to a variety of microscopic models~\cite{mukherjee2025hydrodynamics,saha2023current,bhutia2026diffusive,prakash2025exact,thompson2011lattice}. We define current variables $J_i^{\pm}\left(\tau_j\right)$ as 
\begin{eqnarray}
J_i^{\pm}\left(\tau_j\right)=f_i^{\pm}\left(\tau_{j+1}\right)-f_i^{\pm}\left(\tau_{j}\right).
\end{eqnarray}
The variable $J_i^{\pm}\left(\tau_j\right)$ gives the change in the density of $A$ particles at the site $i$ during a microscopic time step $d\tau=\tau_{j+1}-\tau_j$.
The probability of observing a trajectory of frequency values can be written in terms of these current variables as
\begin{equation}
\begin{aligned}
P\left(\{f^+,f^-\}\right)
&= \Bigg\langle
\prod_{i=-L/2+1}^{L/2}
\prod_{j=1}^{M}
\delta \Big(
f_i^+\left(\tau_{j+1}\right)
- f_i^+\left(\tau_j\right)
- J_i^+\left(\tau_j\right)
\Big) \\
&\quad \times
\delta \Big(
f_i^-\left(\tau_{j+1}\right)
- f_i^-\left(\tau_j\right)
- J_i^-\left(\tau_j\right)
\Big)
\Bigg\rangle_{\{J\}} .
\end{aligned}
\label{prob_traj}
\end{equation}
where $\{f^{\pm}\}=\{f_i^{\pm}(\tau_j),~ i=-L/2+1,...,L/2 \text{ and } j=1,...,M \}$ and $\{J\}=\{J_i(\tau_j),~ i=-L/2+1,...,L/2 \text{ and } j=1,...,M \}$. Also, $\langle\rangle$ denotes an ensemble average over configurations with all possible values of $J_i^{\pm}(\tau_j)$. We next use the continuum representation of the Kronecker delta function. 
$
    \delta \left(a-b\right)=\int_{-\mathrm{i}\pi}^{\mathrm{i}\pi}dk  ~e^{- k\left(a-b\right)}/({2\pi \mathrm{i}})
$
to rewrite Eq.~\eqref{prob_traj} as
\begin{eqnarray}
P\left(\{f^+,f^-\}\right)&=&\langle  \prod_{i=-L/2+1}^{L/2}\prod_{j=1}^{M}\int_{-\mathrm{i} \pi}^{\mathrm{i}\pi}d\hat{f}_i^+\left(\tau_j\right) d\hat{f}_i^-\left(\tau_j\right) ~e^{- \hat{f}_i^+\left(\tau_j\right)\left(f_i^+\left(\tau_{j+1}\right)-f_i^+\left(\tau_j\right)-J_i^+\left(\tau_j\right)\right)}\nonumber \\&\times & e^{- \hat{f}_i^-\left(\tau_j\right)\left(f_i^-\left(\tau_{j+1}\right)-f_i^-\left(\tau_j\right)-J_i^-\left(\tau_j\right)\right)}\rangle_{\{J\}} 
\label{eq:prob_traj}
\end{eqnarray}
\begin{equation}
\begin{aligned}
P\left(\{f^+,f^-\}\right)
&=\int \prod_{j=1}^{M}
\Bigg[
\prod_{i=-L/2+1}^{L/2}
\Bigg\{
d\hat{f}_i^+\left(\tau_j\right) d\hat{f}_i^-\left(\tau_j\right) 
e^{ \Big[
-\hat{f}_i^+\left(\tau_j\right)\big(
f_i^+\left(\tau_{j+1}\right)
- f_i^+\left(\tau_j\right)
\big)-\hat{f}_i^-\left(\tau_j\right)
\big(
f_i^-\left(\tau_{j+1}\right)
- f_i^-\left(\tau_j\right)
\big)
\Big]} 
\Bigg\} \\
&\quad \times
\Bigg\langle
\prod_{i=-L/2+1}^{L/2}
e^{\Big[
\hat{f}_i^+\left(\tau_j\right)
J_i^+\left(\tau_j\right)+\hat{f}_i^-\left(\tau_j\right)
J_i^-\left(\tau_j\right)
\Big]}
\Bigg\rangle_{\{J^j\}}
\Bigg].
\end{aligned}
\end{equation}

Here, $\{J^j\}=\{J_i(\tau_j),~ i=-L/2+1,...,L/2\}$. We need to calculate the average value of the quantity inside the angular brackets $\langle\rangle$.
Denoting 
\begin{equation}
g\left(\mathcal{C}\right)=\prod_{i=-L/2+1}^{L/2}e^{\hat{f}_i^+\left(\tau_j\right)J_i^+\left(\tau_j\right)+\hat{f}_i^-\left(\tau_j\right)J_i^-\left(\tau_j\right)},\label{gc_def}
\end{equation}
we have 
\begin{eqnarray}
\langle \prod_{i=-L/2+1}^{L/2} ~e^{\hat{f}_i^+\left(\tau_j\right)J_i^+\left(\tau_j\right)+\hat{f}_i^-\left(\tau_j\right)J_i^-\left(\tau_j\right)}\rangle &=&\sum_{\mathcal{C}\epsilon \{J^j\} }P\left(\mathcal{C}|\{f^j\}\right)g\left(\mathcal{C}\right)\label{eq:gcav},
\end{eqnarray}
where $P\left(\mathcal{C}|\{f^j\}\right)$ is the probability to observe the configuration $\mathcal{C}$ given the frequencies $\{f^j\}=\{f_i(\tau_j),~ i=-L/2+1,...,L/2\}$ at the time instant $\tau_j$. 

Note that Eq.~\eqref{eq:gcav} can be expressed as a summation over all possible configurations that the system can transition to in a microscopic time step. In each time step, there are only two types of events (reaction or diffusion) that can occur, resulting in a nonzero current of either $+1/N$ or $-1/N$. We separate the configurations $\mathcal{C}$ according to the different microscopic moves. We define the subsets $\mathcal{N}_e^j$, $\mathcal{N}_r^a$,$\mathcal{N}_r^j$ of $\{J^j\}$ as 
\begin{eqnarray}
\nonumber
 \mathcal{N}_e^j&=&\{J_i\left(\tau_j\right), 
i\in \{-L/2+1,...,L/2\}\text{ generated by a exchange move}\},\\
\nonumber
 \mathcal{N}_a^j&=&\{J_i\left(\tau_j\right), 
i \in \{-L/2+1,...,L/2\}\text{ generated by a spin flipping move}\}, \\
  \mathcal{N}_r^j&=&\{J_i\left(\tau_j\right), 
i \in \{-L/2+1,...,L/2\}\text{ generated by a reaction move}\} .
\end{eqnarray}
We also define $\mathcal{C}_0$ as the set of configurations with $J_i(\tau_j)=0$, and therefore using Eq.~\eqref{gc_def}, $g\left(\mathcal{C}_0\right)=1$. As the total probability adds to one, we thus have 
\begin{eqnarray}
P\left(\mathcal{C}_0|\{f^j\}\right)&=&1-\sum_{\mathcal{C}\epsilon \mathcal{N}_e}P\left(\mathcal{C}|\{f^j\}\right)-\sum_{\mathcal{C}\epsilon\mathcal{N}_a}P\left(\mathcal{C}|\{f^j\}\right)-\sum_{\mathcal{C}\epsilon\mathcal{N}_r}P\left(\mathcal{C}|\{f^j\}\right).
\label{eq:pc0pc1}
\end{eqnarray}
From Eq.~\eqref{eq:gcav} and Eq.~\eqref{eq:pc0pc1}, we thus obtain 
\begin{eqnarray}
\langle g\left(\mathcal{C}\right)\rangle_{\{J^j\}}&=& g\left(\mathcal{C}_0\right)+\sum_{\mathcal{C}\epsilon \mathcal{N}_e}P\left(\mathcal{C}|\{f^j\}\right)\left(g\left(\mathcal{C}\right)-1\right)+\sum_{\mathcal{C}\epsilon \mathcal{N}_a}P\left(\mathcal{C}|\{f^j\}\right)\left(g\left(\mathcal{C}\right)-1\right)\nonumber \\ 
\nonumber
&+&\sum_{\mathcal{C}\epsilon \mathcal{N}_r}P\left(\mathcal{C}|\{f^j\}\right)\left(g\left(\mathcal{C}\right)-1\right)
\label{eq:gcjav}\\
&=&1+T_e+T_a+T_r,
\label{eq:gav_Td_Tr}
\end{eqnarray}
where 
\begin{eqnarray}
\label{Te}
T_e&=&\sum_{\mathcal{C}\epsilon \mathcal{N}_e}P\left(\mathcal{C}|\{f^j\}\right)\left(g\left(\mathcal{C}\right)-1\right),\\ 
\label{Ta}
T_a&=& \sum_{\mathcal{C}\epsilon \mathcal{N}_a}P\left(\mathcal{C}|\{f^j\}\right)\left(g\left(\mathcal{C}\right)-1\right),\\
\label{Tr}
T_r&=& \sum_{\mathcal{C}\epsilon \mathcal{N}_r}P\left(\mathcal{C}|\{f^j\}\right)\left(g\left(\mathcal{C}\right)-1\right).
\end{eqnarray}
The quantities $T_e$, $T_a$, and $T_r$ are related to the configuration, which has an exchange move, flipping move, and reaction move, respectively. All the terms are proportional to the microscopic time increments $d\tau$. So Eq.~\eqref{eq:gav_Td_Tr} can be understood as a Taylor expansion of the exponential function up to the leading order as
\begin{eqnarray}
    \langle g\left(\mathcal{C}\right)\rangle_{\{J^j\}}&=& ~e^{T_e+T_a+T_r}+\mathcal{O}\left(d\tau^2\right).
    \label{gc}
\end{eqnarray}
We omit terms of $\mathcal{O}\left(d\tau^2\right)$. The probability of a trajectory of frequencies can now be represented as a path integral~\cite{lefevre2007dynamics},  
\begin{eqnarray}  P\left(\{f^+,f^-\}\right)\!&=&\!\int \prod_{i}\prod_{j}d\hat{f}^+_i\left(\tau_j\right)d\hat{f}^-_i\left(\tau_j\right)~e^{-\mathcal{S}}.  \label{path_integral_dis}
\end{eqnarray}
The exchange moves with a non-zero $\{J^j\}$ correspond to the exchange of $A^{\pm}$ and $B$ particles between nearby sites $i$ and $i+1$. However, reactions occur between $A^{\pm}$ and $B$ particles within the same site. Let us consider each of these terms separately. There are six typical configurations in $\mathcal{N}_e^j$ which correspond to a non-zero $\{J^j\}$.\\
(1)~Exchange of an $B$ particle from site $i$ with a $A^+$ particle from site $i+1$. This happens with probability $R_d d\tau (1-f_i\left(\tau_j\right))f_{i+1}^+\left(\tau_j\right)$ and corresponds to $J_i^+(\tau_j)=1/N$ and $J_{i+1}^+(\tau_j)=-1/N$ and all other $J$s are zero.\\
(2)~Exchange of an $B$ particle from site $i$ with a $A^-$ particle from site $i+1$. This happens with probability $(R_d+\lambda) d\tau (1-f_i\left(\tau_j\right))f_{i+1}^-\left(\tau_j\right)$ and corresponds to $J_i^-(\tau_j)=1/N$ and $J_{i+1}^-(\tau_j)=-1/N$ and all other $J$s are zero.\\
(3)~Exchange of an $A^+$ particle from site $i$ with a $B$ particle from site $i+1$. This happens with probability $(R_d+\lambda) d\tau f_{i}^+\left(\tau_j\right)(1-f_{i+1}\left(\tau_j\right))$ and corresponds to $J_i^+(\tau_j)=-1/N$ and $J_{i+1}^+(\tau_j)=1/N$ and all other $J$s are zero.\\
(4)~Exchange of an $A^+$ particle from site $i$ with a $A^-$ particle from site $i+1$. This happens with probability $R_d d\tau f_{i}^+\left(\tau_j\right)(1-f_{i+1}^-\left(\tau_j\right))$ and corresponds to $J_i^+(\tau_j)=-1/N$ and $J_{i+1}^+(\tau_j)=1/N$ and $J_i^-(\tau_j)=1/N$ and $J_{i+1}^-(\tau_j)=-1/N$ all other $J$s are zero.\\
(5)~Exchange of an $A^-$ particle from site $i$ with a $B$ particle from site $i+1$. This happens with probability $R_d d\tau f_{i}^-\left(\tau_j\right)(1-f_{i+1}\left(\tau_j\right))$ and corresponds $J_{i}^-(\tau_j)=-1/N$ and $J_{i+1}^-(\tau_j)=1/N$ all other $J$s are zero.\\
(6)~Exchange of an $A^-$ particle from site $i$ with a $A^+$ particle from site $i+1$. This happens with probability $R_d d\tau f_{i}^-\left(\tau_j\right)(1-f_{i+1}^+\left(\tau_j\right))$ and corresponds to $J_i^-(\tau_j)=-1/N$ and $J_{i+1}^-(\tau_j)=1/N$ and $J_i^+(\tau_j)=1/N$ and $J_{i+1}^+(\tau_j)=-1/N$ all other $J$s are zero. From all six exchange rules, we have
\begin{eqnarray}
T_e&=&-\sum_{i=-L/2+1}^{L/2}R_d \left(1-f_i\left(\tau_j\right)\right) f^+_{i+1}\left(\tau_j\right)\left(e^{\frac{-\hat{f}^+_{i+1}\left(\tau_j\right)+\hat{f}^+_i\left(\tau_j\right)}{N}}-1\right)\nonumber\\
&+&R_d(1+\lambda) \left(1-f_i\left(\tau_j\right)\right) f^-_{i+1}\left(\tau_j\right)\left(e^{\frac{-\hat{f}^-_{i+1}\left(\tau_j\right)+\hat{f}^-_i\left(\tau_j\right)}{N}}-1\right)\nonumber\\
&+&R_d(1+\lambda)f^+_{i}\left(\tau_j\right) \left(1-f_{i+1}\left(\tau_j\right)\right) \left(e^{\frac{\hat{f}^+_{i+1}\left(\tau_j\right)-\hat{f}^+_i\left(\tau_j\right)}{N}}-1\right)\nonumber\\
&+&R_d f^+_{i}\left(\tau_j\right)f^-_{i+1}\left(\tau_j\right) \left(e^{\frac{\hat{f}^+_{i+1}\left(\tau_j\right)-\hat{f}^+_i\left(\tau_j\right)-\hat{f}^-_{i+1}\left(\tau_j\right)+\hat{f}^-_i\left(\tau_j\right)}{N}}-1\right)\nonumber\\
&+&R_d f^-_{i}\left(\tau_j\right) \left(1-f_{i+1}\left(\tau_j\right)\right) \left(e^{\frac{\hat{f}^-_{i+1}\left(\tau_j\right)-\hat{f}^-_i\left(\tau_j\right)}{N}}-1\right)\nonumber\\
&+&R_d f^-_{i}\left(\tau_j\right)f^+_{i+1}\left(\tau_j\right) \left(e^{\frac{-\hat{f}^+_{i+1}\left(\tau_j\right)+\hat{f}^+_i\left(\tau_j\right)+\hat{f}^-_{i+1}\left(\tau_j\right)-\hat{f}^-_i\left(\tau_j\right)}{N}}-1\right).
\label{Te_new}
\end{eqnarray}
We can have two types of configurations in $\mathcal{N}_a^j$ which correspond to a non-zero $\{J^j\}$.
$A^{\pm}$ particle flips into $A^{\mp}$ with probability $\gamma f_{i}^{\pm}\left(\tau_j\right)$ which correspond  to $J_i^{\pm}(\tau_j)=-1/N$ and $J_i^{\mp}(\tau_j)=1/N$ and all other $J$s are zero. Therefore
\begin{eqnarray}
\label{Ta_new}
T_a=-\sum_i\gamma f^+_{i}\left(\tau_j\right) \left(e^{\frac{-\hat{f}^+_{i}+\hat{f}^-_{i}}{N}}-1\right)-\gamma f^-_{i}\left(\tau_j\right) \left(e^{\frac{-\hat{f}^-_{i}+\hat{f}^+_{i}}{N}}-1\right).
\end{eqnarray}
Similarly, there are four typical configurations in $\mathcal{N}_r^j$ which corresponds to a non-zero $\{J^j\}$.\\
(1)~Reaction of $A^{\pm}$ and $B$ particles at the site $i$ leading to the birth and death of an $A^{\pm}$ and a $B$ particle respectively. This happens with probability $\alpha d\tau f_{i}^{\pm}\left(\tau_j\right)\left(1-f_{i}\left(\tau_j\right)\right)$ and corresponds to $J_i^{\pm}(\tau_j)=1/N$ and all other $J$s are zero.\\
(2)~ The birth and death of a $B$ and an $A^{\pm}$ particle respectively. This happens with probability $\beta d\tau f_{i}^{\pm}\left(\tau_j\right)\left(1-f_{i}\left(\tau_j\right)\right)$ and corresponds to $J_i(\tau_j)=-1/N$ and all other $J$s are zero.\\
Therefore using Eq.~\eqref{Tr}, we obtain
\begin{eqnarray}
\label{Tr_new}
T_r&=&-\sum_i \Bigg[ \alpha d\tau f_{i}^{\pm}\left(\tau_j\right)\left(1-f_{i}\left(\tau_j\right)\right)\left(e^{\frac{\hat{f}_{i}^{\pm}\left(\tau_j\right)}{N}}-1\right)+\beta d\tau f_{i}^{\pm}\left(e^{-\frac{\hat{f}_{i}^{\pm}\left(\tau_j\right)}{N}}-1\right)\Bigg].
\end{eqnarray}
Note that in the above equation, we sum over the site index $i$, since each site is updated during a microscopic step. 
Substituting Eqs.~\eqref{Te_new} \eqref{Ta_new} and \eqref{Tr_new} in Eq.~\eqref{gc} and using this result to compute Eq.~\eqref{eq:prob_traj} yields the path integral in Eq.~\eqref{path_integral_dis} where the action $\mathcal{S}$ in the discrete form is given as 
\begin{eqnarray}
\mathcal{S}&=&\sum_{i=-L/2+1}^{L/2}\sum_{j=1}^M \Bigg[\hat{f}^+_i\left(\tau_j\right)\left(f_i^+\left(\tau_{j+1}\right)-f_i^+\left(\tau_j\right)\right)+\hat{f}^-_i\left(\tau_j\right)\left(f_i^-\left(\tau_{j+1}\right)-f_i^-\left(\tau_j\right)\right)\nonumber\\
&-&R_d \left(1-f_i\left(\tau_j\right)\right) f^+_{i+1}\left(\tau_j\right)\left(e^{\frac{-\hat{f}^+_{i+1}\left(\tau_j\right)+\hat{f}^+_i\left(\tau_j\right)}{N}}-1\right)\nonumber\\
&-&(R_d+\lambda) \left(1-f_i\left(\tau_j\right)\right) f^-_{i+1}\left(\tau_j\right)\left(e^{\frac{-\hat{f}^-_{i+1}\left(\tau_j\right)+\hat{f}^-_i\left(\tau_j\right)}{N}}-1\right)\nonumber\\
&-&(R_d+\lambda)f^+_{i}\left(\tau_j\right) \left(1-f_{i+1}\left(\tau_j\right)\right) \left(e^{\frac{\hat{f}^+_{i+1}\left(\tau_j\right)-\hat{f}^+_i\left(\tau_j\right)}{N}}-1\right)\nonumber\\
&-&R_d f^+_{i}\left(\tau_j\right)f^-_{i+1}\left(\tau_j\right) \left(e^{\frac{\hat{f}^+_{i+1}\left(\tau_j\right)-\hat{f}^+_i\left(\tau_j\right)-\hat{f}^-_{i+1}\left(\tau_j\right)+\hat{f}^-_i\left(\tau_j\right)}{N}}-1\right)\nonumber\\
&-&R_d f^-_{i}\left(\tau_j\right) \left(1-f_{i+1}\left(\tau_j\right)\right) \left(e^{\frac{\hat{f}^-_{i+1}\left(\tau_j\right)-\hat{f}^-_i\left(\tau_j\right)}{N}}-1\right)\nonumber\\
&-&R_d f^-_{i}\left(\tau_j\right)f^+_{i+1}\left(\tau_j\right) \left(e^{\frac{-\hat{f}^+_{i+1}\left(\tau_j\right)+\hat{f}^+_i\left(\tau_j\right)+\hat{f}^-_{i+1}\left(\tau_j\right)-\hat{f}^-_i\left(\tau_j\right)}{N}}-1\right)\nonumber\\
&-&\gamma f^+_{i}\left(\tau_j\right) \left(e^{\frac{-\hat{f}^+_{i}+\hat{f}^-_{i}}{N}}-1\right)-\gamma f^-_{i}\left(\tau_j\right) \left(e^{\frac{-\hat{f}^-_{i}+\hat{f}^+_{i}}{N}}-1\right)\nonumber\\
&-&\alpha d\tau f_{i}^{+}\left(\tau_j\right)\left(1-f_{i}\left(\tau_j\right)\right)\left(e^{\frac{\hat{f}_{i}^{+}\left(\tau_j\right)}{N}}-1\right)-\alpha d\tau f_{i}^{-}\left(\tau_j\right)\left(1-f_{i}\left(\tau_j\right)\right)\left(e^{\frac{\hat{f}_{i}^{-}\left(\tau_j\right)}{N}}-1\right) \nonumber\\
&-&\beta d\tau f_{i}^{+}\left(\tau_j\right)\left(e^{-\frac{\hat{f}_{i}^{+}\left(\tau_j\right)}{N}}-1\right)-\beta d\tau f_{i}^{-}\left(\tau_j\right)\left(e^{-\frac{\hat{f}_{i}^{-}\left(\tau_j\right)}{N}}-1\right)\Bigg].
\label{eq: action_discrete}
\end{eqnarray}
To take the continuum limit, we define the spatial and temporal coordinates $x={ia }$, $t={jd\tau}$, so the summation becomes an integral. We have $\sum_{i=-L/2+1}^{L/2}a \rightarrow \int_{x=-l/2}^{l/2} dx$, $\sum_{j=1}^{M}d\tau \rightarrow \int_{0}^{T=Md\tau} dt$, $f_i^{\pm}(\tau_j) \rightarrow f_i^{\pm}(x,t)$, $\hat{f}^{\pm}_i(\tau_j) \rightarrow \hat{f}_i^{\pm} (x,t)$ and $\prod_{i=-L/2+1}^{L/2}\prod_{j=1}^{M}d\hat{f}_i^{\pm}\left(\tau_j\right) \rightarrow \mathcal{D}\hat{f}^{\pm}(x,t)$. We Taylor expand the following fields up to $\mathcal{O}(a^2)$ and $\mathcal{O}(d\tau)$ to obtain
\begin{eqnarray}
f_i^{\pm}(\tau_{j+1}) &\rightarrow&f^{\pm}(x,t)+{d\tau}\partial_t f^{\pm}(x,t)+\mathcal{O}\left({d\tau^2}\right),\nonumber\\ 
f^{\pm}_{i+1}(\tau_j) &\rightarrow& f^{\pm}(x,t)+a\partial_x f^{\pm}(x,t)+\frac{a^2}{2}\partial_{xx}f^{\pm}(x,t)+\mathcal{O}\left({a^3}\right),\nonumber\\
\hat{f}^{\pm}_{i+1}(\tau_j) &\rightarrow& \hat{f}^{\pm}(x,t)+{a}\partial_x \hat{f}^{\pm}(x,t)+\frac{a^2}{2}\partial_{xx}\hat{f}^{\pm}(x,t)+\mathcal{O}\left({a^3}\right).
\end{eqnarray}
Using the above expansions, Eq.~\eqref{path_integral_dis} now translates to
\begin{eqnarray}
 P\left(\{f^+,f^-\}\right)&=&  \int\mathcal{D}\hat{f}^+(x,t)\hat{f}^-(x,t)~e^{-\mathcal{S}}, \label{path_integral_cont0}
\end{eqnarray}
with the action
\begin{eqnarray}
\mathcal{S}&=&\frac{1}{a}\int_{-l/2}^{l/2}\int_0^{T} dx dt \{\hat{f}^+\partial_{t} f^++\hat{f}^-\partial_{t} f^-+\frac{R_da^2}{N}\partial_xf^+\partial_x\hat{f}^++\frac{R_da^2}{N}\partial_xf^-\partial_x\hat{f}^-\nonumber\\&-&\frac{R_da^2}{N^2}f^+\left(1-f^+\right)\left(\partial_x\hat{f}^+\right)^2-\frac{R_da^2}{N^2}f^-\left(1-f^-\right)\left(\partial_x\hat{f}^-\right)^2+ \frac{2 R_d a^2}{N^2}f^+f^-\partial_x \hat{f}^+\partial_x\hat{f}^-\nonumber\\&-
&\frac{\lambda}{N} a f^+(1-f)\partial_x\hat{f}^++\frac{\lambda}{N} a f^-(1-f)\partial_x\hat{f}^-+\frac{\gamma}{N}\hat{f}^+(f^+-f^-)-\frac{\gamma}{N}\hat{f}^-(f^+-f^-)\nonumber\\
&-&\frac{\alpha}{N} (1-f)\left(f^+\hat{f}^++ f^-\hat{f}^-\right)+\frac{\beta}{N}\left(f^+\hat{f}^++f^-\hat{f}^-\right)\nonumber\\
&-&\frac{1}{2N^2}\left(\alpha(1-f)+\beta \right)\left(f^+(\hat{f}^+)^2+f^-(\hat{f}^-)^2\right)\nonumber\\
&-&\gamma f\frac{(\hat{f}^+-\hat{f}^-)^2}{2N^2}\}.
\label{action_before_scaling}
\end{eqnarray}
In the above expression, we have neglected terms of $\mathcal{O}\left({d\tau^2}\right)$ and $\mathcal{O}\left({a^3}\right)$.  Also, we have neglected terms of $\mathcal{O}\left({N^{-3}}\right)$ in the last term of the above equation.
We introduce the conjugate fields of density and magnetization as $\hat{f}\!=\! (\hat{f}^{+}\! +\! \hat{f}^{-})/2$, $\hat{m}\!=\! (\hat{f}^{+}\!-\! \hat{f}^{-})/2$, respectively.
In terms of these variables, the action takes a simplified form as
\begin{eqnarray}
\mathcal{S}&=&\frac{1}{a}\int_{-l/2}^{l/2}\int_0^{T} dx dt \{\hat{f}\partial_{t} f+\hat{m}\partial_{t}m+\frac{R_da^2}{N}\partial_x f\partial_x\hat{f}\nonumber\\&+&\frac{R_da^2}{N}\partial_x m \partial_x\hat{m}-\frac{R_da^2}{N^2}\{\left(f-f^2\right)\left(\partial_x\hat{f}\right)^2\nonumber\\&+&\left(f-m^2\right)\left(\partial_x\hat{m}\right)^2+2 m\left(1-f\right)\partial_x \hat{f}\partial_x \hat{m}\}\nonumber\\&-
&\frac{\lambda}{N} a m(1-f)\partial_x\hat{f}-\frac{\lambda}{N} a f(1-f)\partial_x\hat{m}+\frac{2\gamma}{N}m\hat{m}\nonumber\\
&-&\frac{s}{N} f \left(1-\frac{f}{Z}\right)\hat{f}-\frac{s}{N}m \left(1-\frac{f}{Z}\right)\hat{m}-\frac{2\gamma}{N^2}f\hat{m}^2\nonumber\\
&-&\frac{1}{2N^2}\left(\alpha(1-f)+\beta \right)\left(f(\hat{f})^2+f(\hat{m})^2+2m\hat{f}\hat{m}\right)\}.\nonumber
\label{action_before_scaling}
\end{eqnarray}
We introduce five new fields $\eta_{d_1}(x,t)$, $\eta_{d_2}(x,t)$, $\eta_f(x,t)$, $\eta_m(x,t)$ and $\eta_a(x,t)$to remove the quadratic terms $\left(\partial_x\hat{f}\right)^2$, $\left(\partial_x\hat{m}\right)^2$, $\hat{f}^2$ and $\hat{m}^2$ respectively and linearize the action through Hubbard-Stratonovich transformations~\cite{pechenik1999interfacial,martin1973statistical} and get the exact FH of the underlying particle model as 
\begin{eqnarray}
\partial_t f&=&D_d\partial_x^2 f-\lambda\partial_x  \left[m(1-f)\right]+s f \left(1-\frac{f}{Z}\right)+\partial_x \eta_{d1}+\eta_f,\nonumber\\
\partial_tm&=&D_d\partial_x^2m-\lambda\partial_x\left[f (1-f)\right]-2 \gamma m+s m \left(1-\frac{f}{Z}\right)+\partial_x \eta_{d2}+\eta_a+\eta_m,
\label{fluctuating_hydro2}
\end{eqnarray}
where the macroscopic parameters $D_d,\lambda,s,Z$ are defined as 
\begin{equation}
D_d=\frac{R_da^2}{N},~~~\lambda:=\frac{\lambda a}{N},~~s=\frac{Z\alpha}{N},~~Z=\frac{\alpha-\beta}{\alpha}.
\label{diff_con}
\end{equation}
Here, $D_d$ and $\lambda$ are the diffusion constants arising from the exchange and the drift constant due to persistence length, respectively. The noiseless version of the above Eq.~\eqref{diff_con} is Eq.~\eqref{eq:ALG_Fisher_Hydro_dimension_less} in the main text.
All the noise terms are Gaussian white noises with mean zero and variance
\begin{eqnarray}
\langle\eta_{d_1} (x_1,t_1)\eta_{d_1} (x_2,t_2)\rangle&=&\frac{2 D_d}{N} f (1-f)\delta (x_1-x_2)\delta (t_1-t_2),\nonumber\\
\langle\eta_{d_2} (x_1,t_1)\eta_{d_2} (x_2,t_2)\rangle&=&\frac{2 D_d}{N}  (f-m^2)\delta (x_1-x_2)\delta (t_1-t_2),
\nonumber\\
\langle\eta_{d_1} (x_1,t_1)\eta_{d_2} (x_2,t_2)\rangle&=&\frac{2D_d}{N} m (1-f)\delta (x_1-x_2)\delta (t_1-t_2),\nonumber\\
\langle\eta_a (x_1,t_1)\eta_a (x_2,t_2)\rangle&=&\frac{2\gamma}{N^2} f
\delta (x_1-x_2)\delta (t_1-t_2),\nonumber\\
\langle\eta_f (x_1,t_1)\eta_f (x_2,t_2)\rangle&=&\frac{2(\alpha+\beta)f-2\alpha f^2}{N^2} \delta (x_1-x_2)\delta (t_1-t_2),
\nonumber\\
\langle\eta_m (x_1,t_1)\eta_m (x_2,t_2)\rangle&=&\frac{2(\alpha+\beta)f-2\alpha f^2}{N^2} \delta (x_1-x_2)\delta (t_1-t_2),
\nonumber\\
\langle\eta_f (x_1,t_1)\eta_m (x_2,t_2)\rangle&=&\frac{2(\alpha+\beta)m-2\alpha m f}{N^2} \delta (x_1-x_2)\delta (t_1-t_2).
\label{correlations2}
\end{eqnarray} 
Eq.~\eqref{fluctuating_hydro2} provides the fluctuating hydrodynamic description of the underlying particle model. There are five distinct noise contributions present in the system. The conservative noises, $\eta_{d_1}$ and $\eta_{d_2}$, arise from diffusive exchanges and appear in both the density and magnetization equations. 
In the scaling regime where diffusion is the fastest process, one may assume local equilibrium, with the equilibrium measure given by a Poisson distribution~\cite{mukherjee2025hydrodynamics,prakash2025exact,bhutia2026diffusive}. In some related models, the hydrodynamics has even been shown to be exact~\cite{kourbane2018exact}. Under such scaling, the conservative noises are significant in the early-time dynamics~\cite{bhutia2026diffusive}, while they become irrelevant in the late-time regime.   We note that the multiplicative noises are fundamentally coupled because birth-death events act on the active species, breaking standard equilibrium fluctuation-dissipation theorems.

The non-conservative noises $\eta_f$ and $\eta_m$ originate from the stochastic birth--death dynamics. An important consequence of these noises can already be understood in zero dimensions: unlike the deterministic hydrodynamic description, where the carrying capacity corresponds to a fixed point, these fluctuations remain finite even when the density reaches carrying capacity. The noise vanishes only when both the density and magnetization go to zero. Such multiplicative noises play an important role in population dynamics, as they can drive the formation of macroscopic domains from initially uniform states and lead to coarsening behavior~\cite{korolev2010genetic,hallatschek2010life,hallatschek2023proliferating}.  
There is an additional noise term, $\eta_a$, arising from the stochastic flipping (tumbling) dynamics. This noise remains active as long as the density is finite. All noise terms are assumed to be Gaussian, consistent with the central limit theorem, since each coarse-grained box contains a large number of lattice sites.
\section{Details of the microscopic simulations}
\label{1d_mc_details}
 A kinetic Monte Carlo algorithm is employed to obtain the evolution of microscopic observables. A lattice with $L$ sites is initialized, where each site is associated with three densities, $A_p$, $A_m$, and $B$, corresponding to the densities of $A^+$, $A^-$, and $B$ particles, respectively. Each of these densities can increase or decrease by a factor of $1/N$ during a Monte Carlo step, where $N$ is the total number of particles on each lattice point. 

 At each Monte Carlo step, a site $i$ is chosen randomly, and a probability-rate vector $\text{Prob}$ consisting of 13 possible events is computed as follows:
\begin{eqnarray}
\nonumber
\text{Prob}(1)=\alpha A_p(i)B(i)\,dt, \qquad
\text{Prob}(2)=\beta A_p(i)\,dt,\\
\text{Prob}(3)=\alpha A_m(i)B(i)\,dt, \qquad
\text{Prob}(4)=\beta A_m(i)\,dt,
\end{eqnarray}
corresponding to four reaction processes:
\begin{eqnarray}
\text{Prob}(5)&=&R_d B(i)A_p(i+1)\,dt,\nonumber\\
\text{Prob}(6)&=&(R_d+\lambda) B(i)A_m(i+1)\,dt,\nonumber\\
\text{Prob}(7)&=&(R_d+\lambda) A_p(i)B(i+1)\,dt,\nonumber\\
\text{Prob}(8)&=&R_d A_p(i)A_m(i+1)\,dt,\nonumber\\
\text{Prob}(9)&=&R_d A_m(i)B(i+1)\,dt,\nonumber\\
\text{Prob}(10)&=&R_d A_m(i)A_p(i+1)\,dt,
\end{eqnarray}
which represent six exchange processes:
\begin{equation}
\text{Prob}(11)=\gamma A_p(i)\,dt, \qquad
\text{Prob}(12)=\gamma A_m(i)\,dt,
\end{equation}
corresponding to two flipping processes.
Finally,
\begin{equation}
\text{Prob}(13)=1-\sum_{j=1}^{12} \text{Prob}(j),
\end{equation}
represents the probability that none of the above processes occur.

At each time step, one of these 13 processes is selected according to the probability rates ,and the density variables $A_p(i)$, $A_m(i)$, and $B(i)$ are updated depending on the selected process. After each update, boundary conditions are imposed to ensure that all densities remain bounded between $0$ and $1$.

\section{Phase diagram}
\label{Finite system size}
The phase diagram shown in Fig.~\ref{fig:profile}$(a)$ in the main text exhibits a band of unstable wavenumbers. Pattern formation occurs when the system size is sufficiently large for at least one of the modes within this unstable band to be accessible. The competition between reaction and activity controls the extent and location of this band: the reaction terms stabilize the homogeneous state, whereas activity promotes its destabilization. Consequently, increasing the reaction rate progressively narrows the unstable band and shifts it toward larger $\Pe$. This behavior is shown in Fig.~\ref{fig:phase_diagram_system_size}.

The long-wavelength instability is suppressed at zero reaction rate in a finite system unless the activity is sufficiently large. For finite reaction strength $s>0$, the homogeneous mode remains stable, since
\begin{equation}
\Lambda^+(0)=-s\qquad (s<2),
\end{equation}
so the instability cannot originate from the $k=0$ mode. Instead, activity can destabilize finite-wavelength modes. Expanding the positive eigenvalue for small $k$ gives
\begin{equation}
\Lambda^+(k)=-s+\left[-1+\frac{\text{Pe}^2(2Z-1)(1-Z)}{2-s}\right]k^2+O(k^4),
\label{eq:eig_small_k}
\end{equation}
for $s<2$. Thus, although the homogeneous mode remains stable, sufficiently strong activity can render long-wavelength modes unstable.

In a finite system of linear size $L$, the smallest nonzero accessible wavenumber is
\begin{equation}
k_{\min}=\frac{2\pi}{L}.
\end{equation}
Consequently, the onset of the instability is determined by the growth rate of this lowest Fourier mode,
\begin{equation}
\Lambda^+(k_{\min})=-s+\left[-1+\frac{\text{Pe}^2(2Z-1)(1-Z)}{2-s}\right] \left(\frac{2\pi}{L}\right)^2+\mathcal{O}(L^{-4}).
\label{eq:eig_small_k_L}
\end{equation}
The lowest accessible mode therefore becomes unstable when the activity exceeds a finite-size-dependent threshold, even though the homogeneous mode remains linearly stable.

\begin{figure}
\includegraphics[width=1\linewidth]{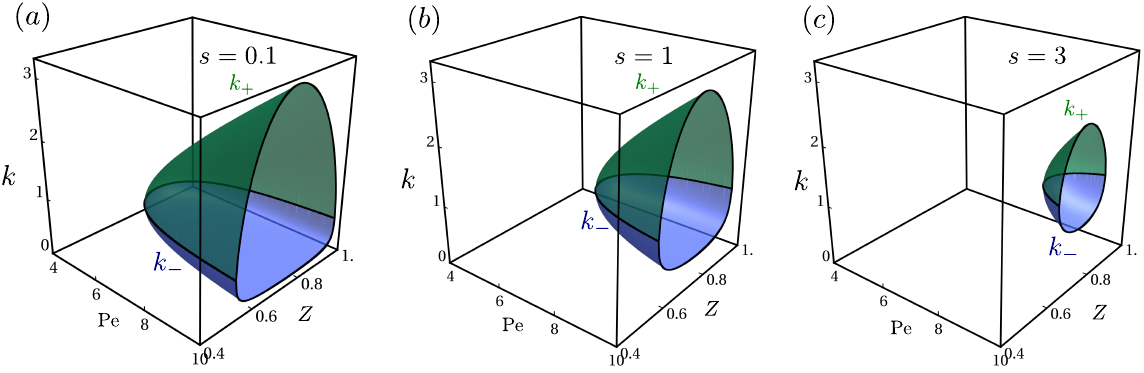}
\caption{The unstable band progressively narrows and shifts toward larger $\Pe$ as the reaction rate increases. $k_+$ and $k_-$} represent the maximum and minimum values of the unstable mode, respectively.
\label{fig:phase_diagram_system_size}
\end{figure}

\section {Steady-state length scale in one dimension}
\label{kmax_1d}
The steady state of Eq.~\eqref{eq:hydro2} in the main text is a periodic solution when the parameter lies inside the unstable phase of Fig.~\ref{fig:profile}($a$) in the main text. It has multiple domains. The length scales of these domains can be obtained from the calculation of their structure factor as follows: 
\begin{eqnarray} K_{max}=\frac{\sum_{k} k |\hat{f}(k)|^2}{\sum_{k} |\hat{f}(k)|^2}. 
\label{eq:k_max_str}\end{eqnarray} 
The value of $K_{max}$ is close to the maximum unstable mode shown in (Fig.~\ref{fig:profile}($c$) in the main text). The exact value of the maximum unstable mode is determined as $\Lambda^+$. \begin{eqnarray} \Lambda^+_{max}=\frac{ \sqrt{ Z^{2}\alpha^{2} -4Z\alpha\gamma +4\gamma^{2} -\lambda^{4} +6Z\lambda^{4} -13Z^{2}\lambda^{4} +12Z^{3}\lambda^{4} -4Z^{4}\lambda^{4} } }{ \lambda\sqrt{4-12Z+8Z^{2}} } .
\label{eq:lambda_max}
\end{eqnarray}
\begin{figure}[h]
    \centering    \includegraphics[width=0.9\linewidth]{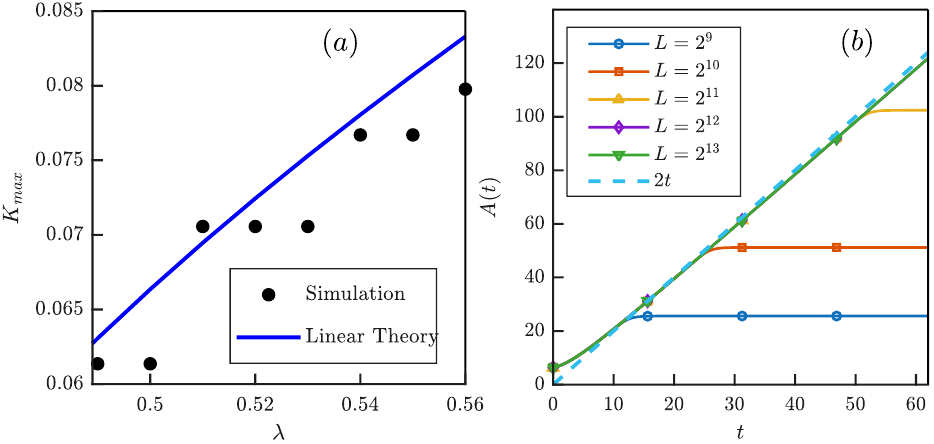}
    \caption{$(a)$ The dependence of $K_{\rm max}$ on $\lambda$ for $\gamma = 0.01$, $Z = 0.8$, and $\bar{\alpha} = 0.09$ is plotted. The system size is chosen as $L = 2^{10}$. The theoretical curve is obtained from the linear stability analysis based on Eq.~\eqref{eq:lambda_max}. The simulation data are obtained by numerically integrating the full non-linear hydrodynamic equations, Eq.~\eqref{eq:hydro2} in the main text, and measuring the steady-state properties after evolving the system up to $t = 2^{20}$. $(b)$ The area under the curve vs. time for different system sizes is plotted by solving the non-linear hydrodynamic equations Eq.~\eqref{eq:hydro2} in the main text 
using a pseudo-spectral method. A first-order Euler time integration scheme is used with $\delta t=2^{-5}$. The system parameters are chosen as follows:  $N=1000$, $R_d=1$, $\lambda=0$, $\gamma=0.01 $, $\alpha=0.01, Z=1$.
}
\label{fig:Kmax_vs_Pe}
\end{figure}
The discrepancy between the characteristic length scales obtained from Eq.~\eqref{eq:k_max_str} and Eq.~\eqref{eq:lambda_max} at high activity is a consequence of the non-linear coupling among the Fourier modes within the unstable band, shown in Fig.~\ref{fig:profile}($c$) in the main text. On the other hand, the theoretical prediction can be tested in the regime where the unstable band is narrow. Fig.~\ref{fig:Kmax_vs_Pe}($a$) compares the steady-state length scale obtained from the full non-linear equation with that predicted by the linear theory, Eq.~\eqref{eq:lambda_max}. The linear theory captures the qualitative behavior of the steady-state length scale.
\section{Dominant balance at the tip of magnetization front profile}
\label{quasi-static}
We begin with the continuum equations governing the density $f(x,t)$ and the magnetization $m(x,t)$,
\begin{eqnarray}
\partial_t f &=& \,\partial_x^2 f
-\Pe \partial_x m(1-f) 
+s f(1-f/Z),\nonumber\\
\partial_t m &=& \,\partial_x^2 m
-\Pe \partial_x f(1-f) 
-2 m
+s (1-f/Z)m.
\end{eqnarray}

To obtain an approximate analytical description of the propagating front, we employ a dominant-balance analysis of the magnetization field. Numerical solutions of both the continuum equations show that the magnetization remains small throughout the front. Deep inside the bulk, the relaxation term $-2\gamma m$ rapidly suppresses the magnetization, while the symmetric production of positive and negative particles leads to a negligible net contribution from the birth-death dynamics. Near the leading edge of the front, the numerical results (Fig.~\ref{fig:dominant_balance}) reveal that the time derivative of the magnetization balances the diffusive transport and reaction terms, while the two active contributions, proportional to $\lambda$ and $\gamma$, balance each other separately.
\begin{figure}[h]
    \centering
    \includegraphics[width=0.4\linewidth]{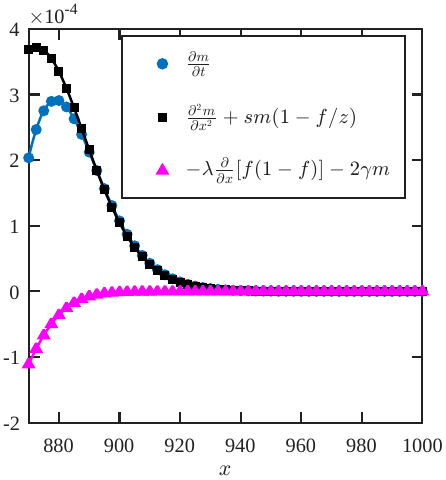}
    \caption{Near the tip of the magnetization profile, the time derivative is balanced by the combined diffusive and reaction contributions, while the two active terms, proportional to $\lambda$ and $\gamma$, balance each other. The system parameters are $L=2^{10},\lambda=0.05,\gamma=0.01,\alpha=0.02$. The profile is shown at time $t=2^{7}$.
}
    \label{fig:dominant_balance}
\end{figure}

Under these assumptions, the magnetization becomes locally slaved to the density field, yielding
\begin{equation}
m(x,t)
=
-\frac{\Pe}{2}
\partial_x\!\left[f(1-f)\right].
\label{eq:approx}
\end{equation}
Substituting this expression into the density equation leads to a closed non-linear evolution equation for the density,
\begin{equation}
\partial_t f
=
\,\partial_x^2 f
+\kappa(4f-3)(\partial_x f)^2
+\kappa(1-f)(1-2f)\partial_x^2 f
+s f(1-f/Z),
\end{equation}
where $\kappa= \Pe^2/2.
$

This equation represents an effective scalar field theory obtained by integrating out the rapidly relaxing magnetization field. Unlike previously studied scalar active matter models, such as the active Model $A/B/C$ family or the scalar theories proposed for motility-induced phase separation~\cite{cates2019active,cates2025active}, the activity here generates both a density-dependent effective diffusivity and a non-linear gradient-squared term. These features place the resulting equation in a distinct class of scalar active field theories. It can also be noted that there is no effective diffusion constant.
\section{Traveling waves in the presence of activity}
\label{active_wave}
In the presence of weak activity, the qualitative features of these traveling waves remain largely unchanged, although the propagation speed increases compared to the passive case. However, when the activity becomes sufficiently strong, or when the system parameters lie within the unstable region of [Fig.~\ref{fig:profile}($a$) in the main text], the wave structure becomes distorted and deviates significantly from the  FKPP behavior. The steady-state velocity of a traveling wave can be calculated by computing the area under the frequency vs. distance
curve as $A(t)=\int_x f(x,t)~dx$. $A(t)$  increases with time and finally takes a saturation value due to the size effect. The steady-state velocity is calculated before the wave is affected by the boundary. We calculate the velocity with different system sizes as shown in Fig.~\ref{fig:Kmax_vs_Pe}($b$). For a sufficiently large system size, a linear fit of the area vs. time plot gives the velocity of the traveling waves.
\begin{figure}
    \centering    
    \includegraphics[width=0.9\linewidth]{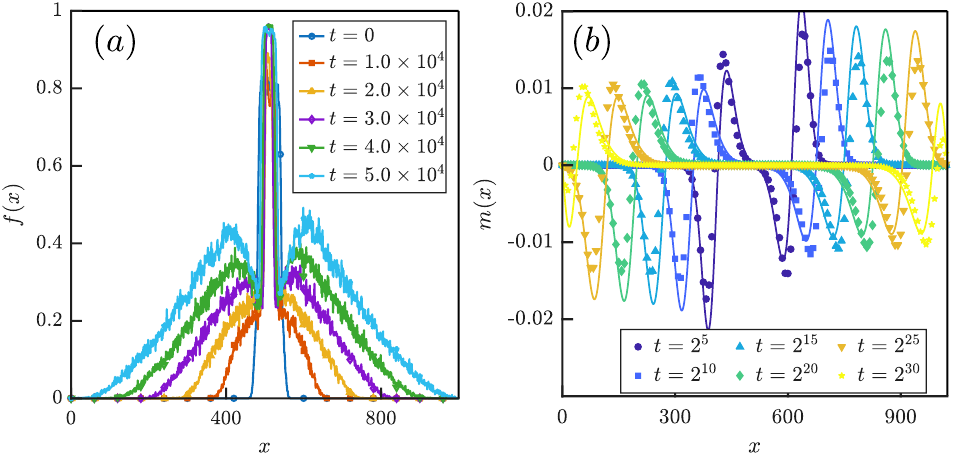}
    \caption{(a) Time evolution of the density profile in the high-activity regime obtained from Monte Carlo simulations. The system parameters are $R_d=0.5$, $\lambda=1$, $\gamma=0.01$, and $\alpha=0.01$. The initial magnetization profile is uniform and equal to zero throughout the system. (b)The solid line represents the numerical solution of Eq.~\eqref{eq:hydro2} in the main text for $\lambda=0.05$, $\gamma=0.02$, and $R_d=1$, while the symbols correspond to the approximate solution obtained from Eq.~\eqref{eq:approx}. The approximation remains accurate over a broad range of times.}    \label{fig:magnetization_wave}
\end{figure}
On the tip of the wave, the frequency and magnetization field are both very small (see Fig.~\ref{fig:magnetization_wave}($a$) and Fig.~\ref{fig:profile}($f$) in the main text), so ignoring the quadratic terms, Eq.~\eqref{eq:ALG_Fisher_Hydro_dimension_less} becomes 
\begin{eqnarray}
\partial_t f&=&\Delta f-\Pe  \nabla m + \bar{s}f,\nonumber\\
\partial_t m&=&\Delta m -\text{Pe} \nabla f-2 m+ \bar{s} m .
\label{eq:ALG_Fisher_Hydro_linear}
\end{eqnarray}
In a frame that moves with constant velocity $v$, both $f(x,t)$ and $m(x,t)$ now become $\rho(u)$ and $m(u)$, where $u=x-vt$. So the above equation becomes 
\begin{eqnarray}
-v f'&=& f''-\text{Pe} ~m' + s f,\nonumber\\
-v m'&=& m'' -\text{Pe} ~f'-2 m+s m .
\label{eq:ALG_Fisher_Hydro_linear_u_frame}
\end{eqnarray}
Substituting the derivative of the magnetization yields
\begin{eqnarray}
-v f'&=& f''+ s f+\frac{\text{Pe}}{v}(m'' -\text{Pe} ~f'-2 m+ s m ).
\end{eqnarray}
Assuming that the wave has an exponentially decaying front of type $e^{-l u}$, the above equation gives
\begin{eqnarray}
v l&=& l^2+s+\frac{\text{Pe}}{v}(l^2+l~\Pe -2 + s).
\end{eqnarray}
From the above equation, the velocity of the traveling wave becomes 
\begin{eqnarray}
v=\frac{(l^2+s)+\sqrt{(l^2+s)^2+4 l \text{Pe} (s-2+l^2+l \text{Pe} )}}{2 l}.
\label{eq:vel_sel_Pe}
\end{eqnarray}
In the absence of activity ($\text{Pe}=0$) the above equation gives $v=\frac{(l^2+s)}{ l}$.
The minimum of this convex function is $V_{\min}=2\sqrt{s}$. For sufficiently localized initial conditions, the asymptotic front dynamics selects this minimal velocity~\cite{van1988front}. 
Fig.~\ref{fig:profile}($f$) in the main text shows that the agreement between the linear theory and the numerical solution of the full non-linear equations deviates as the P\'eclet number increases. 
This departure may indicate that non-linear effects in the bulk become increasingly important and contribute to the selection of the wave speed. This is in contrast to pulled waves, for which the propagation speed is determined solely by the linear dynamics at the leading edge of the front.

Fig.~\ref{fig:magnetization_wave}($a$) shows that the nature of the propagating wave changes substantially in the high-activity regime. In particular, the wave is no longer a pulled front, as indicated by the distortion of the density profile in the bulk. The enhanced density at the center results from the actively driven particles from both directions. 
This departure from the wave-like solution at high activity can be understood from the phase portrait of Eq.~\eqref{eq:position_momentum}. At low activity [Fig.~\ref{fig:stream_Pe0_Pe10}($a$)], the phase portrait exhibits streamlines connecting the unstable fixed point $(1,0)$ to the stable fixed point $(0,0)$, corresponding to the wave-like solution. In contrast, at high activity [Fig.~\ref{fig:stream_Pe0_Pe10}($b$)], a singularity develops in the momentum field, causing the wave-like trajectory to terminate at the singularity rather than reach the stable fixed point.

\begin{figure}[h]
\centering
\includegraphics[width=0.5\linewidth]{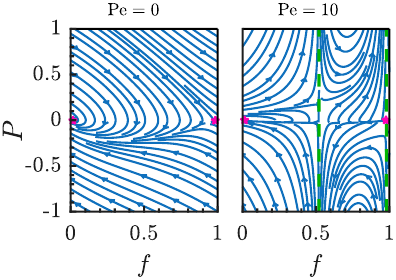}
\caption{Phase portraits of Eq.~\eqref{eq:position_momentum} in the main text for $\Pe=0$ and $\Pe=10$. At $\Pe=0$, the streamline connects the unstable fixed point $(1,0)$ to the stable fixed point $(0,0)$. At $\Pe=10$, a singularity develops in the momentum variable, indicated by the dotted lines.}
\label{fig:stream_Pe0_Pe10}
\end{figure}
\section{Breakdown of Hydrodynamics}
\label{breakdown_hydro}
The hydrodynamic theory is expected to be quantitatively accurate when the bias parameter $\lambda$ is small compared to the diffusion rate. Throughout the analysis presented in Figs.~\ref{fig:profile}($b$),($c$),($e$),($f$) in the main text, this condition is satisfied, leading to excellent agreement between the hydrodynamic theory and the Monte Carlo simulations. However, when the bias becomes comparable to the diffusion, deviations from the hydrodynamic predictions emerge. This is evident in Fig.~\ref{fig:microphase_th_sim}, where the discrepancies between the theoretical and simulation results arise because $\lambda=0.5$ and $R_d=1$ are of the same order, whereas the exact hydrodynamics limit corresponds to $\frac{\lambda}{R_d} \ll 1$. Nevertheless, the hydrodynamic theory correctly captures the emergence of the micro-phase-separated state, demonstrating that it successfully predicts the underlying instability responsible for micro-phase formation.
\begin{figure}[h]
    \centering
\includegraphics[width=0.4\linewidth]{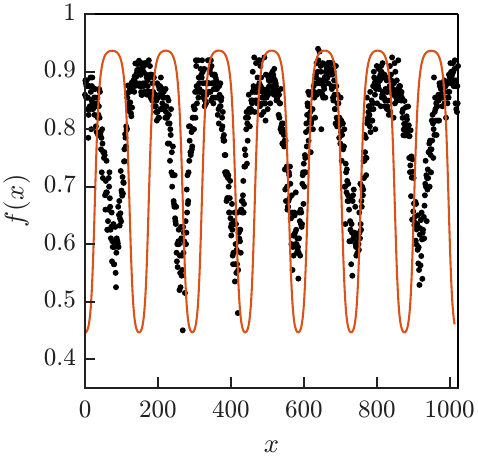}
    \caption{The formation of microphase separation. The points represent the results obtained from Monte Carlo simulations, while the solid lines correspond to the predictions of the hydrodynamic theory. The parameters used are $\lambda=0.5$, $\gamma=0.01$, $\bar{\alpha}=0.01$, $L=2^{10}$, and $N=100$. The system is initialized with a non-uniform density, and the steady-state profile is plotted at time $T=2^{20}$. The value of $\lambda/R_d=0.5$, which is same to Fig.~\ref{fig:profile}$(c)$ in the main text.}
\label{fig:microphase_th_sim}
\end{figure}

\section{The active stepping-stone model in two dimensions}
The one-dimensional model can be naturally extended to two dimensions. In 2D, active tumbling dynamics is modeled using four discrete polarization states, indexed by $a \in \{0, 1, 2, 3\}$, which correspond to biased motion along the $+x$, $+y$, $-x$, and $-y$ directions, respectively~\cite{jose2022active}. Active biased hopping occurs with rate $\lambda$ and is subject to local crowding, requiring the nearest-neighbor target site to be vacant. Concurrently, the isotropic diffusive exchange process, with rate $R_d$, occurs identically among all species. In addition to transport, particles undergo local on-site reproduction and death governed by the same reaction rules introduced in Fig.~\ref{fig: SS_Model} in the main text. From these microscopic rules, the hydrodynamic equations for the density of each individual polarization component $f^a(\mathbf{r}, t)$ can be derived analogous to the one-dimensional case. The resulting hydrodynamic equations exhibit excellent agreement with particle-level simulations, even in the regime where active bias and diffusive transport are of comparable magnitude. 

\label{sec:ASS_2d}
\begin{figure}[t!]
    \centering    \includegraphics[width=1\linewidth]{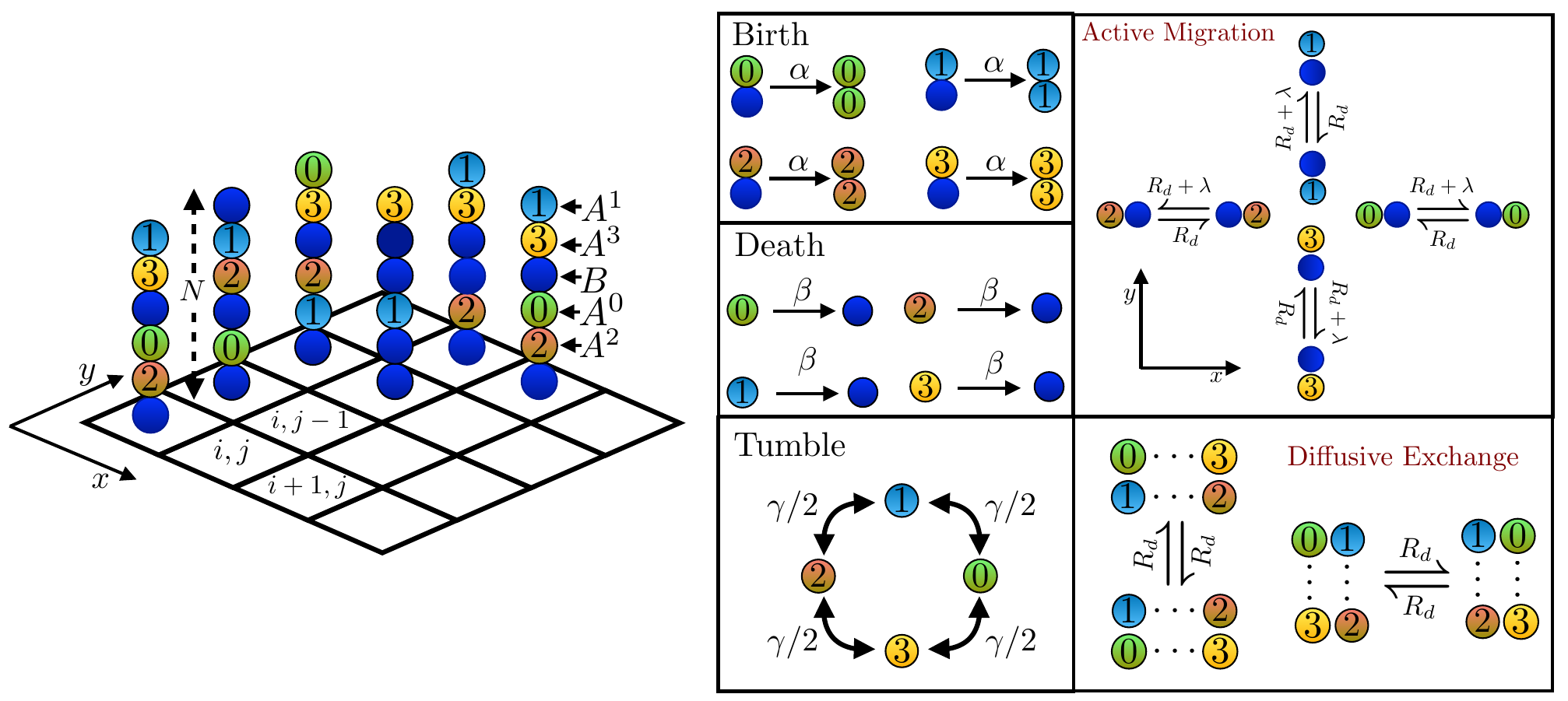}
    \caption{A schematic illustration of a two-dimensional run-and-tumble particle is shown, 
possessing four possible spin states: $m \in \{0, 1, 2, 3\}$, corresponding 
to motion that is biased along the positive $x$,  positive $y$, negative $x$, and negative $y$ 
directions, respectively. Transitions between adjacent polarity states occur at a rate 
$\gamma/2$, as indicated by the arrows in the figure. In addition to the biased motion corresponding to active migration, the particles $A^{(0)-(3)}$ can also perform diffusive exchange with any of the other $A^{(0)-(3)}$ particles along the $x$ or $y$ directions. The birth-death rules for each type of particle remain the same as the one-dimensional case.}  
\label{fig:2d_rtp_flip_rule}
\end{figure}
From these microscopic dynamics, the hydrodynamic equations for the density of each component can be derived as
\begin{eqnarray}
\partial_t f_0&=&D_s\nabla^2 f_0-\lambda\nabla_x f_0 (1-f)+\frac{\gamma}{2}(f_1+f_3-2f_0)+Z\alpha f_0\left(1-\frac{f}{Z}\right),\nonumber\\
\partial_t f_1&=&D_s\nabla^2 f_1-\lambda\nabla_y f_1 (1-f)+\frac{\gamma}{2}(f_2+f_0-2f_1)+Z\alpha f_1\left(1-\frac{f}{Z}\right),\nonumber\\
\partial_t f_2&=&D_s\nabla^2 f_2+\lambda\nabla_x f_2 (1-f)+\frac{\gamma}{2}(f_1+f_3-2f_2)+Z\alpha f_2\left(1-\frac{f}{Z}\right),\nonumber\\
\partial_t f_3&=&D_s\nabla^2 f_3+\lambda\nabla_y f_3 (1-f)+\frac{\gamma}{2}(f_2+f_0-2f_3)+Z\alpha f_3\left(1-\frac{f}{Z}\right),
\end{eqnarray}
where $f=f_0+f_1+f_2+f_3$ is the total density of $A$ type of particle, $\nabla^2=\partial_x^2+\partial_y^2$,$\nabla_x=\partial_x$,$\nabla_y=\partial_y$.
Following a similar procedure to the one-dimensional case, we can derive a hydrodynamic equation in two dimensions analogous to Eq.~\eqref{eq:hydro2} in the main text, where the magnetization field becomes a two-dimensional vector. 
\begin{eqnarray}
\partial_t f&=&D_s \nabla^2 f-\lambda \vec{\nabla}. \left[ \vec{m} (1-f)\right]+Z\alpha  f\left(1-\frac{f}{Z}\right),\nonumber\\
\partial_t \vec{m}&=&D_s\nabla^2 \vec{m} -\lambda \vec{\nabla}. \left[f (1-f)\right]-2 \gamma \vec{m}+Z\alpha  \left(1-\frac{f}{Z}\right)\vec{m}.
\label{eq:ALG_Fisher_2D}
\end{eqnarray}
This is Eq.~\eqref{eq:hydro2} in the main text. The resulting system exhibits qualitatively similar behavior to its one-dimensional counterpart, including the emergence of MIPS and their modification in the presence of reactions. The hydrodynamic theory is verified with the particle-based simulation. It shows the signature of both motility and proliferation. The full evolution equation is shown in Fig.~\ref{fig:2d_sim_fd_time}.

\begin{figure}[h]
    \centering
\includegraphics[width=0.6\linewidth]{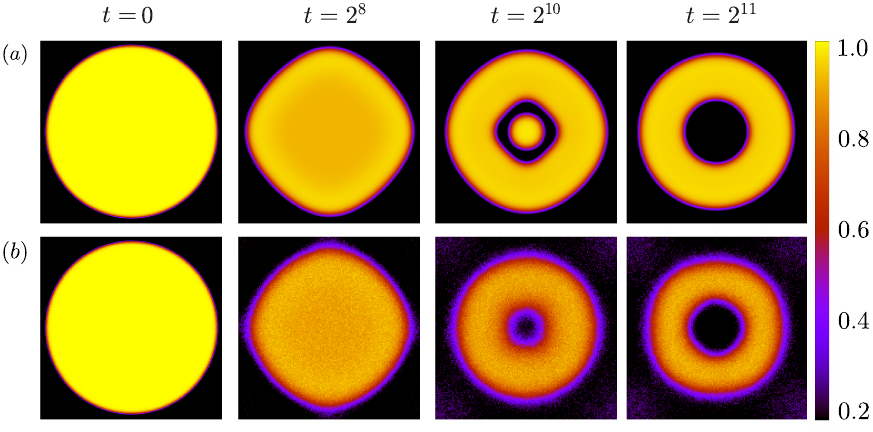}
    \caption{Numerical verification of the hydrodynamic equation in two dimensions is presented. (a) shows the temporal evolution obtained by solving Eq.~\eqref{eq:hydro2} in the main text using a pseudo-spectral method, while (b) shows the corresponding particle-level Monte Carlo simulations. The two characteristic time scales, corresponding to the activity-dominated and reaction-dominated regimes, are shown at $t=2^8$ and $t=2^{11}$, respectively.
}
\label{fig:2d_sim_fd_time}
\end{figure}

Two-dimensional particle simulations in the hydrodynamic limit are computationally expensive, particularly when the bias parameters are orders of magnitude smaller compared to the diffusion coefficient. In this regime, the system requires prohibitively long simulation times to reach the steady state and develop the distinct morphologies. 
\begin{figure}[h]
    \centering \includegraphics[width=0.5\linewidth]{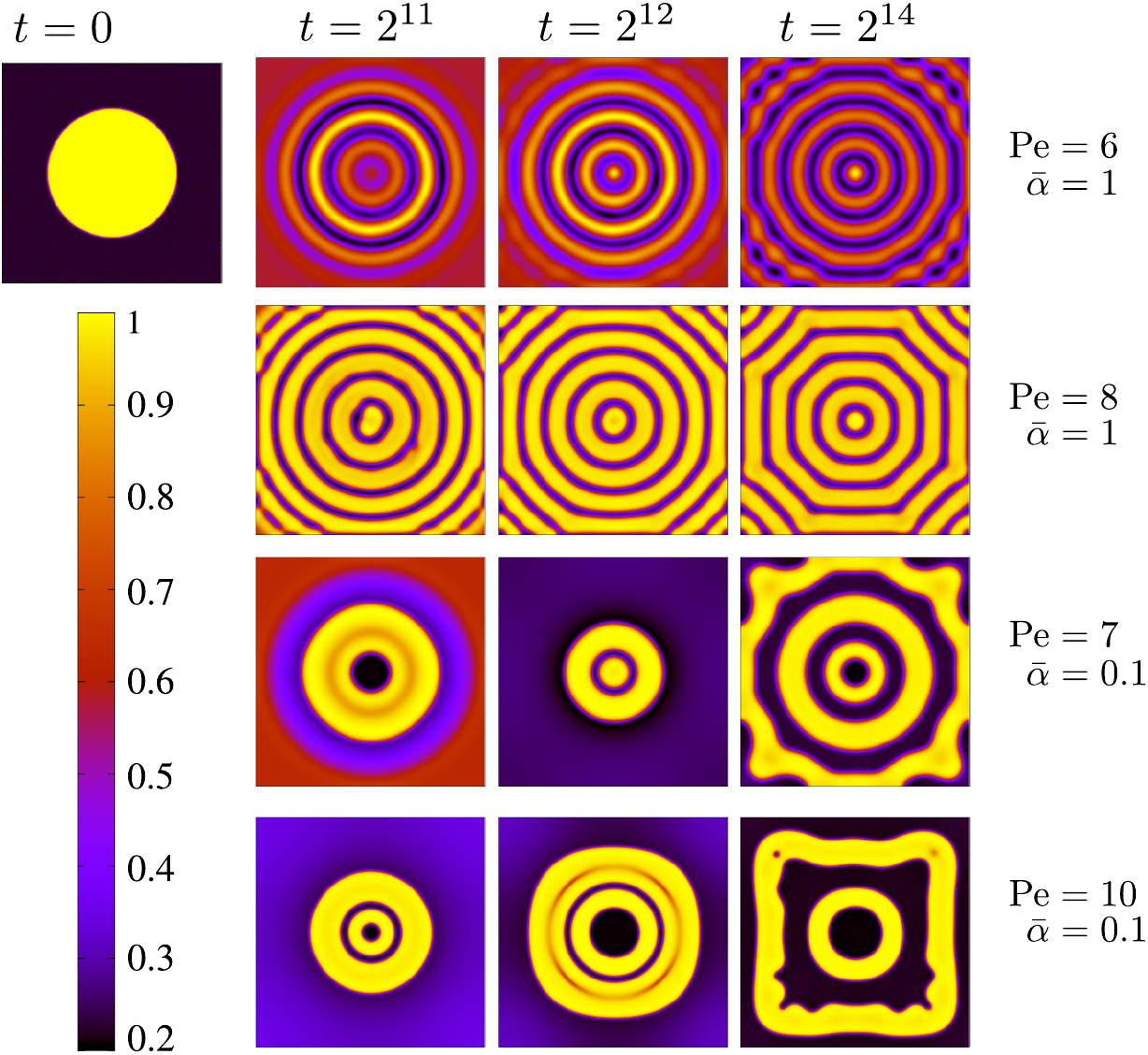}
    \caption{The density fields as functions of $x$ and $y$ are obtained by numerically solving Eq.~\eqref{eq:ALG_Fisher_2D} using a pseudo-spectral method. The system is initialized in a non-homogeneous state with random magnetization with mean zero and variance $0.5$. The time evolution is carried out using a forward Euler scheme until the system reaches a steady state. For all the figures, the parameters are chosen as $R_d=1,\gamma=0.01, L=256$. For the top panel, the parameters are set to $\lambda = 0.6$ and $\alpha = 0.01$, while for the panel below, $\lambda = 0.8$ and $\alpha = 0.01$ are used. For the bottom-most panel, $\lambda = 0.7$ and $\alpha = 0.001$ are taken, whereas for the panel immediately above it, $\lambda = 1$ and $\alpha = 0.001$ are used.}
    \label{fig:ring_patterns}
\end{figure}
Therefore, instead of performing particle-level simulations, we evolve the full non-linear hydrodynamic equation, Eq.~\eqref{eq:hydro2} in the main text, to obtain the different steady-state morphologies, as shown in Fig.~\ref{fig:ring_patterns}. This confirms that the model has a rich morphological behaviour with respect to non-uniform initial conditions.
\subsection{Steady state of the two-dimensional active stepping stone model}
\label{steady state}
The Figs.~\ref{fig:2d_sim_th_plots}($c$)-($f$) in the main text show the system in its steady state. To confirm this, we calculate the average and variance of the density as a function of time. Under periodic boundary conditions, Eq.~\eqref{eq:hydro2} in the main text yields
\begin{eqnarray}
\partial_t \langle f\rangle=s\langle f\rangle-s\langle f^2\rangle/Z.
\end{eqnarray}
Here, $\langle O\rangle=\frac{1}{A}\int_A O(x,y)dxdy$ denotes the spatial average of an observable $O$, where $A$ is the area of the system. The diffusive and active terms are conservative and therefore vanish upon spatial averaging under periodic boundary conditions. So in the steady state we have $\langle f\rangle=\langle f^2\rangle/Z$. Fig.~\ref{fig:steady_state_2d_main} shows that the system has reached steady state. The average density differs significantly from the initial density; this is because of the non-conservative reaction term.

\begin{figure}[t!]
    \centering   \includegraphics[width=0.9\linewidth]{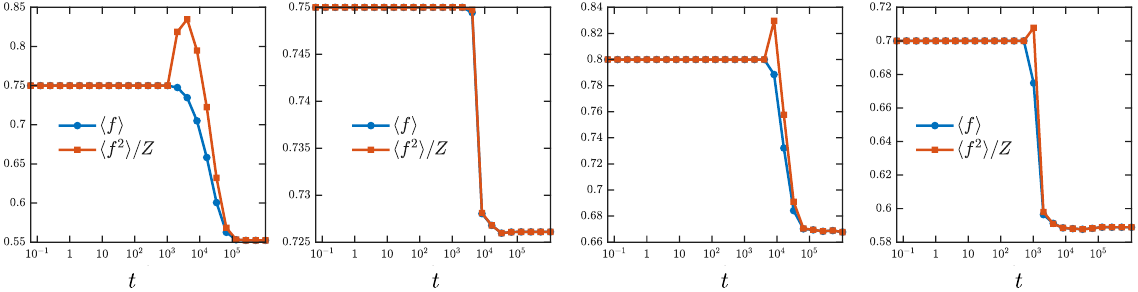}
    \caption{The average density and its variance are plotted as functions of time for (Figs.~\ref{fig:2d_sim_th_plots}($c$)-($f$) in the main text), respectively. The system is considered to have reached a steady state when $\langle f\rangle=\langle f^2\rangle/Z$.
}    \label{fig:steady_state_2d_main}
\end{figure}
\clearpage
\end{widetext}
\end{document}